\pdfoutput=1 
\documentclass[%
reprint,
superscriptaddress,
 amsmath,amssymb,
 aps,
prb,
floatfix,
twocolumn
]{revtex4-2}

\usepackage{physics}
\usepackage{graphicx}
\usepackage{dcolumn}
\usepackage{bm}
\usepackage[svgnames,dvipsnames]{xcolor}
\usepackage[colorlinks,citecolor=DarkGreen,linkcolor=magenta,linktocpage]{hyperref}

\makeatletter
\let\MYcaption\@makecaption
\makeatother

\usepackage[labelformat=simple]{subcaption}
\makeatletter
\let\@makecaption\MYcaption
\makeatother

\begin{document}

\title{Quantum many-body scars of spinless fermions with density-assisted hopping in higher dimensions}

\author{Kensuke Tamura}
    \email{tamura-kensuke265@g.ecc.u-tokyo.ac.jp}
    \affiliation{Department of Physics, Graduate School of Science, The University of Tokyo, 7-3-1 Hongo, Tokyo 113-0033, Japan}

\author{Hosho Katsura}
    \affiliation{Department of Physics, Graduate School of Science, The University of Tokyo, 7-3-1 Hongo, Tokyo 113-0033, Japan}
    \affiliation{Institute for Physics of Intelligence, The University of Tokyo, 7-3-1 Hongo, Tokyo 113-0033, Japan}
    \affiliation{Trans-scale Quantum Science Institute, The University of Tokyo, Bunkyo-ku, Tokyo 113-0033, Japan}

\date{\today}

\begin{abstract}
We introduce a class of spinless fermion models that exhibit quantum many-body scars (QMBS) originating from kinetic constraints in the form of density-assisted hopping. 
The models can be defined on any lattice in any dimension and allow for spatially varying interactions. 
We construct a tower of exact eigenstates with finite energy density, and we demonstrate that these QMBS are responsible for the nonthermal nature of the system by studying the entanglement entropy and correlation functions.
The quench dynamics from certain initial states is also investigated, and it is confirmed that the QMBS induce nonthermalizing dynamics.
As another characterization of the QMBS, we give a parent Hamiltonian for which the QMBS are unique ground states. We also prove the uniqueness rigorously. 
\end{abstract}

\maketitle

\section{\label{sec:introduction}Introduction}
The dynamics of isolated quantum systems has long been of great interest from a fundamental point of view~\cite{polkovnikov2011colloquium,nandkishore2015many}. 
In recent years, it has gained further interest with the improvement of experimental techniques in ultracold atomic systems. 
Generic isolated systems are expected to eventually thermalize regardless of the initial state, meaning that macroscopic quantities relax to their equilibrium values described by statistical mechanics.
Theoretically, the thermalization mechanism relies on the eigenstate thermalization hypothesis (ETH)~\cite{deutsch1991quantum,srednicki1994chaos,rigol2008thermalization}, a strong version of which asserts that all energy eigenstates are thermal, i.e., they are locally indistinguishable from the microcanonical ensemble with the corresponding energy.
While the ETH has been confirmed for many interacting quantum systems~\cite{rigol2008thermalization,kim2014testing,d2016quantum}, there are exceptions.
For example, 
integrable and many-body localized systems fail to thermalize due to exact or emergent conservation laws. 
\par
Recent experiments using Rydberg atoms~\cite{bernien2017probing} and quantum simulations in tilted optical lattices~\cite{su2022observation} observed long-lived coherent oscillations of local observables for particular initial states.
This is an unexpected behavior given the nonintegrability of the system, and it offers another example of a quantum many-body system that evades thermalization.
This nonthermalizing dynamics is attributed to the existence of atypical eigenstates dubbed quantum many-body scars (QMBS)~\cite{serbyn2021quantum,regnault2022quantum,chandran2022quantum,papic2021weak}, which have nonthermal properties, and hence violate the ETH. 
These QMBS have a sub-volume-law entanglement scaling even though they are in the middle of the energy spectrum. 
The above experiment has led to a surge of interest in understanding the nonthermal behavior of QMBS. 
In addition to investigating effective models relevant to the experiments~\cite{choi2019emergent,ho2019periodic,lin2019exact,khemani2019signatures,iadecola2019quantum, desaules2022weak}, 
there have been a number of attempts to construct nonintegrable models with exact QMBS. 
To date, a plethora of methods to construct models with exact QMBS have been developed~\cite{shiraishi2017systematic,moudgalya2018entanglement,pai2019dynamical,chattopadhyay2020quantum,moudgalya2020quantum,pakrouski2020many,moudgalya2020large,iadecola2020quantum,kuno2020flat,sugiura2021many,langlett2022rainbow}.
However, the overwhelming majority of previous studies have focused on one-dimensional systems, although some examples in two or more dimensions are known~\cite{schecter2019weak,ok2019topological,lee2020exact,surace2020weak,michailidis2020stabilizing,lin2020quantum,wildeboer2021topological,PhysRevB.102.224303}.
For a deeper understanding of the mechanism of QMBS, it is helpful to have a variety of models with exact QMBS in higher dimensions; therefore, new examples of such models are desired.
Furthermore, the majority of the studies deal with spin systems, whereas there are fewer examples of QMBS in particle systems such as fermionic systems~\cite{vafek2017entanglement,mark2020eta,moudgalya2020eta,hart2020compact,desaules2021proposal,pakrouski2021group,yoshida2022exact,nakagawa2022exact}.
\par
In this paper, we construct and study a class of spinless fermionic models with QMBS in higher dimensions by exploiting a kinetic constraint known as density-assisted hopping or density-induced tunneling~\cite{dutta2015non,ruhman2017topological,gotta2021two-prl,gotta2021two-prb}.
This kind of interaction appears in effective models of time-periodic systems~\cite{di2014quantum} and systems with large tilted potential, and it can be realized in ultracold atomic systems~\cite{kohlert2021experimental}.
In addition, it has been studied in the context of superconductivity as well~\cite{crepel2022unconventional}.
Previous studies have identified a class of kinetically constrained models that exhibit QMBS and scarred dynamics for bosons~\cite{hudomal2020quantum, zhao2020quantum} and fermions~\cite{zhao2021orthogonal} in one dimension. 
In contrast, our method is capable of constructing models in arbitrary higher dimensions. 
Further, the models obtained this way are not necessarily translation invariant and thus can be regarded as disordered models.
A few examples of QMBS coexisting with disorder are known~\cite{shibata2020onsager, van2021disorder}, and our construction provides examples of disordered quantum many-body scarred models. 
As another characterization of QMBS, we also give a parent Hamiltonian for which the QMBS constructed are unique ground states. 
\par
The paper is organized as follows.
In Sec.~\ref{sec:model}, we define our model in arbitrary dimensions.
We verify the nonintegrability of the model through the level spacing statistics.
In Sec.~\ref{sec:qmbs}, we first show how to construct the exact QMBS in the system. 
We then demonstrate that the QMBS are nonthermal by studying the entanglement entropy and correlation functions.
We also argue that the dynamics of entanglement entropy and fidelity exhibit a nonthermal nature. 
In Sec.~\ref{sec:The parent Hamiltonian}, we show that it is possible to construct a parent Hamiltonian such that the QMBS are its unique ground states. 
We conclude with a summary in Sec.~\ref{sec:summary}. 
In Appendix~\ref{appendix:particle-hole transformation}, we introduce another form of the Hamiltonian with QMBS related to the original one by a particle-hole transformation.
A generalization of our construction of the scarred model is shown in Appendix~\ref{appendix:generalization}.
In Appendix~\ref{appendix:Q operator and pairing states}, we see that the operator that generates the exact QMBS can be regarded as a spinless analog of the $\eta$-pairing operator.
Finally, in Appendix~\ref{appendix:Proof of the uniqueness}, we present a rigorous proof that the QMBS are the unique ground states of the parent Hamiltonian.
\section{\label{sec:model}Model}
\subsection{\label{subsec:hamiltonian} Hamiltonian}
\begin{figure}[tb]
\centering
\includegraphics[width=0.8\columnwidth]{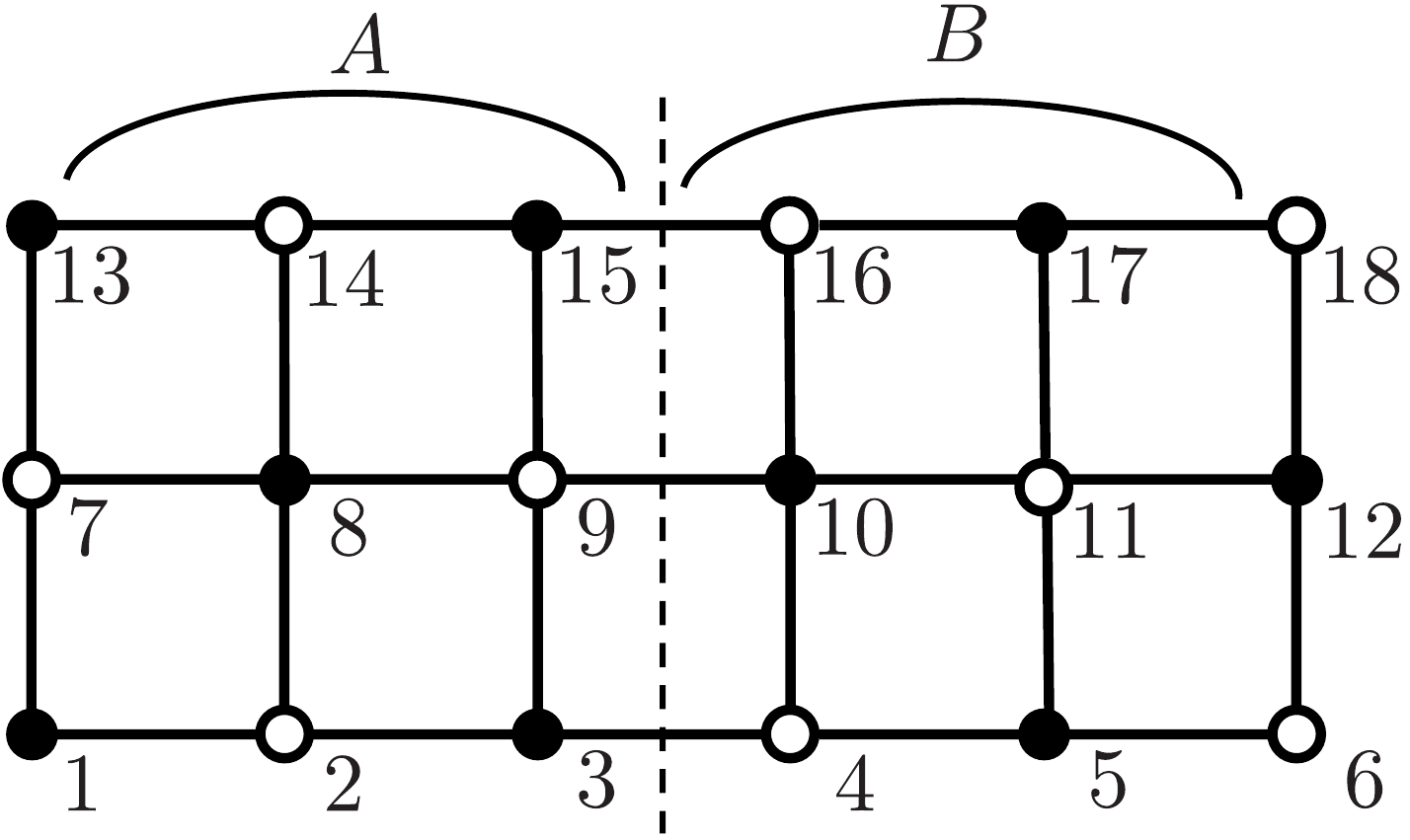}
\caption{An example of a bipartite lattice.
The black and white sites belong to sublattices $\Lambda^{(1)}$ and $\Lambda^{(2)}$, respectively.
Regions $A$ and $B$ are defined for the calculation of the
entanglement entropy.
The integers are labels for the lattice sites.
}
\label{fig:two-dim lattice}
\end{figure}

We consider a system of spinless fermions on a lattice $(\Lambda, \mathcal{B})$, where $\Lambda$ is the set of sites, and $\mathcal{B}$ is the set of bonds.
An element of $\mathcal{B}$ is a pair of different sites, $\langle x, y \rangle$, and we identify $\langle x, y\rangle$ with $\langle y, x\rangle$.
Here, we assume that the lattice is bipartite, i.e., its sites can be partitioned into two disjoint sets $\Lambda^{(1)}$ and $\Lambda^{(2)}$, and for any $\langle x, y\rangle \in \mathcal{B}$, one has either $x \in \Lambda^{(1)}$, $y \in \Lambda^{(2)}$ or $x \in \Lambda^{(2)}$, $y \in \Lambda^{(1)}$ (see, e.g., Fig.~\ref{fig:two-dim lattice} and ignore the dashed line and symbols $A$ and $B$ for now).
We denote by $c_x^\dag$ and $c_x$, respectively, the creation and the annihilation operators at site $x \in \Lambda$.
They satisfy 
\begin{align}
    \{c_x, c_y\} &= \{c_x^\dag, c_y^\dag\} = 0,  \label{eq:anticommutation relation} \\
    \{c_x, c_y^\dag\} &= \delta_{x, y}, \label{eq:anticommutation relation 2}
\end{align}
for $x, y \in \Lambda$.
In the following, $\mathcal{V}$ denotes the whole Fock space spanned by states of the form $\prod_{x \in \Lambda}\left(c_x^\dag\right)^{n_x} \ket{\mathrm{vac}} (n_x = 0, 1)$, where $\ket{\mathrm{vac}}$ is the vacuum state annihilated by all $c_x$. 
The Hamiltonian is given by
\begin{align}
    H &= H_{\mathrm{hop}} + H_{\mathrm{cor}}, \label{eq:hamiltonian}\\
    H_{\mathrm{hop}} &= \sum_{x, y \in \Lambda} t_{x, y} c_x^\dag c_y,\label{eq:hopping_term}\\
    H_{\mathrm{cor}} &= \sum_{x \in \Lambda} A_x \left(\sum_{y \in \Lambda} q_{x, y} c_y^\dag \right)c_x c_x^\dag \left(\sum_{y' \in \Lambda} q_{x, y'} c_{y'}\right). \label{eq:correlated_hopping_term}
\end{align}
We assume that the hopping matrix $\mathsf{T} = (t_{x, y})_{x, y \in \Lambda}$ is real symmetric and that $t_{x, y}$ can be nonzero only when $\langle x, y\rangle \in \mathcal{B}$.
We also assume that $\mathsf{Q} = \left(q_{x,y}\right)_{x, y \in \Lambda}$ is a real skew-symmetric matrix of the form 
\begin{align}
    q_{x, y} = 
    \begin{cases}
        t_{x, y} \ \ \text{for} \ x \in \Lambda^{(1)} \ \text{and}\  y \in \Lambda^{(2)},  \\
        -t_{x, y} \ \ \text{for} \ x \in \Lambda^{(2)} \ \text{and}\  y \in \Lambda^{(1)}, \\ 
        0 \ \ \text{otherwise}.
    \end{cases}
\end{align}
The coefficients $A_x$ are arbitrary real numbers.
We note in passing that another form of the Hamiltonian~\eqref{eq:hamiltonian} is obtained by a particle-hole transformation.
See Appendix~\ref{appendix:particle-hole transformation} for details.
\par 
Let us make some comments on the model.
First, the model has the density-assisted hopping term described by Eq.~\eqref{eq:correlated_hopping_term}.
To illustrate the term, let us consider a periodic chain with $L$ sites, where $L$ is even so that the lattice is bipartite.
For simplicity, here we consider the case in which the hopping matrix is translation invariant, i.e., the matrix elements of $\mathsf{T}$ is  $t_{j, j+1} = t_{j+1, j} = t$ for all $j = 1, \dots, L$, although the coefficients $A_j$ may be site-dependent.
Then, the density-assisted hopping term takes the following form: 
\begin{align}
    H_{\mathrm{cor}}
    & = t^2  \sum_{j=1}^L A_j \left(
    n_{j-1} + n_{j+1} - n_{j-1}n_j - n_j n_{j+1} \right. \nonumber \\ 
    & \left.+ c_{j-1}^\dag c_{j+1}  + c_{j+1}^\dag c_{j-1} \right. \nonumber \\
    &\left. - c_{j-1}^\dag n_j c_{j+1} - c_{j+1}^\dag n_j c_{j-1}
    \right), \label{eq:one_dimensional_correlated_hopping}
\end{align}
where $n_j = c_j^\dag c_j$.
The model without the terms in the first row of Eq.~\eqref{eq:one_dimensional_correlated_hopping} has been discussed in the literature~\cite{bariev1991integrable,chhajlany2016hidden}.
\par
Second, our model can be defined on an arbitrary bipartite lattice in any dimension.
Examples include two-dimensional square and three-dimensional cubic lattices with open boundary conditions.
In addition, our construction can be extended to more general lattices that are not necessarily bipartite.
See Appendix~\ref{appendix:generalization} for the general construction.
\par
Finally, we mention that our model in one dimension is related to the spin model discussed in~\cite{shibata2020onsager}.
In fact, we can map the one-dimensional model \eqref{eq:one_dimensional_correlated_hopping} to a spin model using the Jordan-Wigner transformation defined by
\begin{align}
    c_j = \left(\prod_{k=1}^{j-1}(-\sigma_k^z)\right)\sigma_j^{+},
\end{align}
where $(\sigma_j^x, \sigma_j^y, \sigma_j^z)$ are the Pauli matrices at site $j$ and $\sigma_j^{+} = (\sigma_j^x + i \sigma_j^y)/2$. 
With this transformation, the Hamiltonian~\eqref{eq:one_dimensional_correlated_hopping} is mapped to the following one for a spin-chain:
\begin{align}
    &H_{\mathrm{cor}} 
    = \sum_{j=2}^{L-1} 
    \left\{ 2 t^2 A_j \dyad{010}{010} \right. \nonumber \\
    &~ \left.+ t^2 A_j \left(\ket{011} + \ket{110} \right) \left(\bra{011} + \bra{110}\right)
    \right\}_{j-1, j, j+1}+ H_{\rm bdy}, \label{eq:spin model}
\end{align}
where the states $\ket{0}$ and $\ket{1}$ are eigenstates of $\sigma^z_j$ with eigenvalues $+1$ and $-1$, respectively. 
The last term denoted by $H_{\rm bdy}$ in Eq.~\eqref{eq:spin model} is a non-local operator arising from the periodic boundary conditions.
Ignoring this term, Eq.~\eqref{eq:spin model} coincides with the perturbation term $H_{{\rm pert},n}$ discussed in~\cite{shibata2020onsager} with $n=2$, $c_j^{(1)} = c_j^{(2)} = 2t^2 A_j$, and $c_j^{(3)} = 0$~\footnote{The term corresponding to $c_j^{(3)}$ gives rise to a nonlocal term that does not commute with the fermionic parity, rendering its fermionic counterpart unphysical.}.
Since in one dimension the hopping Hamiltonian $H_{\rm hop}$ is equivalent to the $S=1/2$ XX chain, the entire Hamiltonian $H = H_{\rm hop} + H_{\rm cor}$ maps to the scarred model discussed in~\cite{shibata2020onsager}.   
In this sense, our construction can be regarded as an extension of the particular case of the one-dimensional model to a wider class of lattice fermion systems.
\subsection{\label{subsec:non-integrability} Nonintegrability of the model}
\begin{figure}[tb]
\includegraphics[width=0.8\columnwidth]{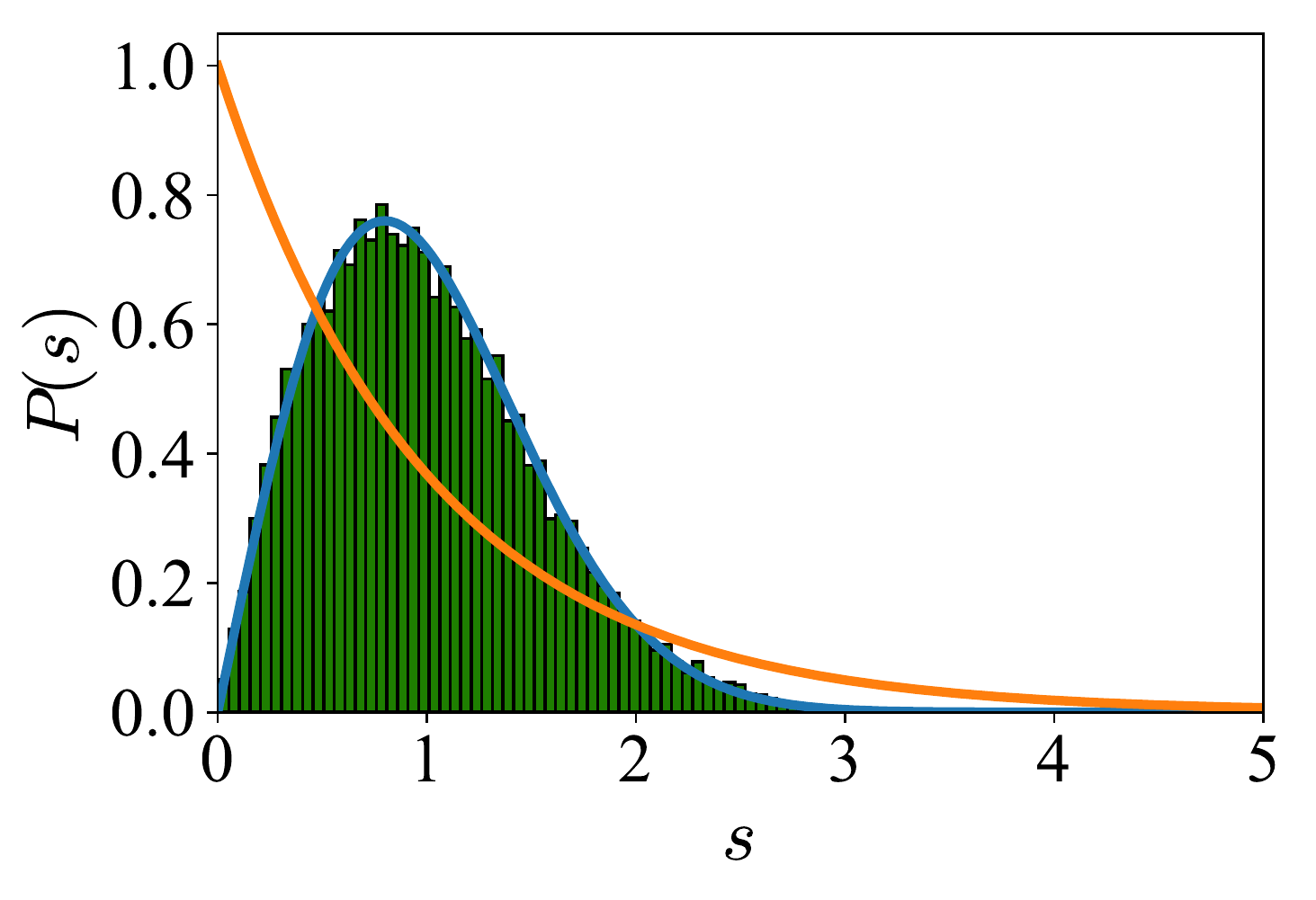}
\caption{The level statistics in the middle half of the spectrum of $H$ on a $6 \times 3$ lattice $(N=8)$ with open boundary conditions as shown in~Fig.\ref{fig:two-dim lattice}.
The matrix elements of $\mathsf{T}$ and $A_x$ are chosen uniformly at random from $[0.5, 1.5]$ and $[-0.5, 0.5]$, respectively.
The Wigner-Dyson (blue) and the Poisson (orange) distributions are shown for comparison. 
} \label{fig:level_spacing}
\end{figure}

For the model to be a nontrivial example of a scarred model, it must be nonintegrable. 
To verify this, we study the level statistics of the model.
It is performed in a sector with a fixed number of particles since the Hamiltonian~\eqref{eq:hamiltonian} conserves the total number of particles $N = \sum_{x \in \Lambda} n_x$, where $n_x = c_x^\dag c_x$.
Let $\{E_i\}_{i=1, 2, \dots}$ be a set of energy eigenvalues in ascending order, and let $s_i = (E_{i+1} - E_i)/\Delta$ be the level spacing normalized by the mean level spacing $\Delta$. 
For nonintegrable systems with time-reversal symmetry, the level-spacing distribution is given by the Wigner-Dyson distribution $P(s) = (\pi/2)s  e^{-\pi s^2/4}$, whereas for integrable systems it is given by the Poisson distribution $P(s) = e^{-s}$~\cite{casati1985energy,prosen1993energy}. 
We show the distribution of the level spacings of our model in Fig.~\ref{fig:level_spacing}.
The histogram confirms that the distribution is well described by the Wigner-Dyson distribution.
We also compute the level-spacing ratio $r_i$~\cite{pal2010many} defined by $r_i  =  \min(s_i, s_{i+1})/\max(s_i, s_{i+1})$, which characterizes the distribution quantitatively.
It is known that the average value of $r_i$ is $\langle r \rangle \simeq 0.53590$ for the Wigner-Dyson distribution and $\langle r \rangle \simeq 0.38629$ for the Poisson distribution~\cite{atas2013distribution}.
The average value of $r_i$ calculated from the histogram in Fig.~\ref{fig:level_spacing} is $0.52927...$, which is close to that of the Wigner-Dyson distribution.
Therefore, we can conclude that our model is nonintegrable.
\section{\label{sec:qmbs}Quantum many-body scarred states}
In this section, we construct a series of exact eigenstates of the Hamiltonian $H$ in Eq.~\eqref{eq:hamiltonian}.
We provide evidence that these eigenstates are QMBS of the system. 

\subsection{\label{subsec:qmbs eigenstate} Definition of exact QMBS}
Here, we show how to construct a series of exact eigenstates of $H$ algebraically. 
To this end, we define an operator $Q$ by 
\begin{align}
    Q = \sum_{x, y \in \Lambda} q_{x, y} c_x c_y, \label{eq:q operator}
\end{align}
and consider the following states
\begin{align}
    \ket{\Psi_k} = Q^k \ket{\overline{\mathrm{vac}}}, k = 0, 1, .\dots, \left\lfloor \frac{|\Lambda|}{2} \right\rfloor, \label{eq:qmbs}
\end{align}
where $\ket{\overline{\mathrm{vac}}} = \prod_{x \in \Lambda} c_x^\dag \ket{\mathrm{vac}}$ and $\lfloor x \rfloor$ denotes the floor function~\footnote{Here and throughout the paper, $|{\cal S}|$ denotes the number of elements in a set ${\cal S}$.}.
Since $Q$ is a sum of operators each of which annihilates two fermions, $\ket{\Psi_k}$ is a state with $(|\Lambda| - 2k)$ particles, where the integer $k$ is at most $\lfloor |\Lambda|/2 \rfloor$.
We now show that the states $\ket{\Psi_k}$ are zero-energy eigenstates of $H_{\mathrm{hop}}$ and $H_{\mathrm{cor}}$, and hence $H \ket{\Psi_k} = 0$. 
\par
First, one finds 
\begin{align}
    [H_{\mathrm{hop}}, Q] = -2 \sum_{x, x' \in \Lambda} \left(\mathsf{Q} \mathsf{T}\right)_{x, x'} c_x c_{x'}=0. \label{eq:com_rel_hop_q}
\end{align}
This is because the matrix $\mathsf{Q} \mathsf{T}$ is real symmetric and $c_x c_{x'}$ is antisymmetric under the exchange of $x$ and $x'$.  
Hence, the operator $Q$ commutes with $H_{\mathrm{hop}}$.
Noting that $H_{\mathrm{hop}}\ket{\mathrm{\overline{vac}}} = 0$, we have
\begin{align}
    H_{\mathrm{hop}} \ket{\Psi_k} = 0. \label{eq:hhop=0}
\end{align}
The commutation relation~\eqref{eq:com_rel_hop_q} also follows from the fact that the operator $Q$ is a sum of pair operators, each of which commutes with $H_{\mathrm{hop}}$. This can be seen by noting that each pair consists of two eigenmodes of $H_{\mathrm{hop}}$ with opposite energies. 
(See Appendix~\ref{appendix:Q operator and pairing states} for a detailed discussion.)
In this sense, the operator $Q$ can be thought of as a spinless analog of Yang's $\eta$-pairing operator~\cite{yang1989eta}, as discussed recently for a class of one-dimensional models in Ref.~\cite{gotta2022exact}. 
\par
Next, we see that $H_{\mathrm{cor}}$ also annihilates the state $\ket{\Psi_k}$. 
Here, we note that the operator $Q$ does not commute with $H_{\mathrm{cor}}$, hence $Q$ is not a conserved quantity of the Hamiltonian $H$.
Instead of $H_{\mathrm{cor}}$ itself, we compute the commutation relation between $c_x^\dag \left(\sum_{y \in \Lambda} q_{x, y} c_y\right)$ and $Q$, and we find 
\begin{align}
    \left[ \, c_x^\dag \left(\sum_{y \in \Lambda} q_{x, y} c_y\right), Q \, \right] =
    2 \qty( \sum_{y \in \Lambda} q_{x, y} c_y )^2 = 0,
\end{align}
for all $x \in \Lambda$.
Since the diagonal elements of $\mathsf{Q}$ are zero, we see that $c_x^\dag \left(\sum_{y \in \Lambda} q_{x, y} c_y\right) \ket{\overline{\mathrm{vac}}} = 0$ and that $c_x^\dag \left(\sum_{y \in \Lambda} q_{x, y} c_y\right) \ket{\Psi_k} = 0$. 
Thus we have
\begin{align}
    H_{\mathrm{cor}} \ket{\Psi_k} = 0.
\end{align}
This result, together with \eqref{eq:hhop=0}, yields 
\begin{align}
    H \ket{\Psi_k} = 0,
\end{align}
which is the desired result. 
\par
We remark that the Hamiltonian $H$ and the operator $Q$ satisfy a restricted spectrum generating algebra~\cite{buvca2019non, moudgalya2020eta}. 
Here we use the notation of Ref.~\cite{moudgalya2020eta}. 
A Hamiltonian $H$ is said to exhibit a restricted spectrum generating algebra of order 1 (RSGA-1) if there exist a state $\ket{\psi_0}$ and an operator $\eta^\dag$ such that $\eta^\dag \ket{\psi_0} \neq 0$, and they satisfy
\begin{align}
    & \textrm{(i)}\ H\ket{\psi_0} = E_0 \ket{\psi_0}, \\
    & \textrm{(ii)}\ [\,H, \eta^\dag\,]\ket{\psi_0} = \mathcal{E} \eta^\dag \ket{\psi_0}, \\
    & \textrm{(iii)}\ [\,[\,H, \eta^\dag\,], \eta^\dag\,] = 0.
\end{align}
In our case, we have $H \ket{\overline{\mathrm{vac}}} = 0$ and can prove $[\,H, Q\,] \ket{\overline{\mathrm{vac}}} = 0$ by a similar calculation as above.
We can also see that $[\,[\,H, Q\,], Q\,] = 0$ by repeating the same calculations. 
Therefore, the Hamiltonian~\eqref{eq:hamiltonian} exhibits RSGA-1 with $\ket{\psi_0} = \ket{\overline{\mathrm{vac}}}$, $\eta^\dag = Q$, $E_0 = 0$, and $\mathcal{E} = 0$. 
The zero-energy states generated by repeated applications of $Q$ on $\ket{\overline{\mathrm{vac}}}$, i.e., Eq. \eqref{eq:qmbs}, can be thought of as QMBS associated with this algebra. 

\subsection{\label{subsec:properties} Violation of the ETH}
Here we demonstrate that the states~\eqref{eq:qmbs} are nonthermal by examining the entanglement entropy and expectation values of a physical quantity.
\begin{figure}[tb]
    \centering
    \begin{tabular}{c}
        \begin{minipage}[h]{\linewidth}
            \centering
            \subcaption{\label{fig:energy_vs_ee}}
            \includegraphics[width=\linewidth]{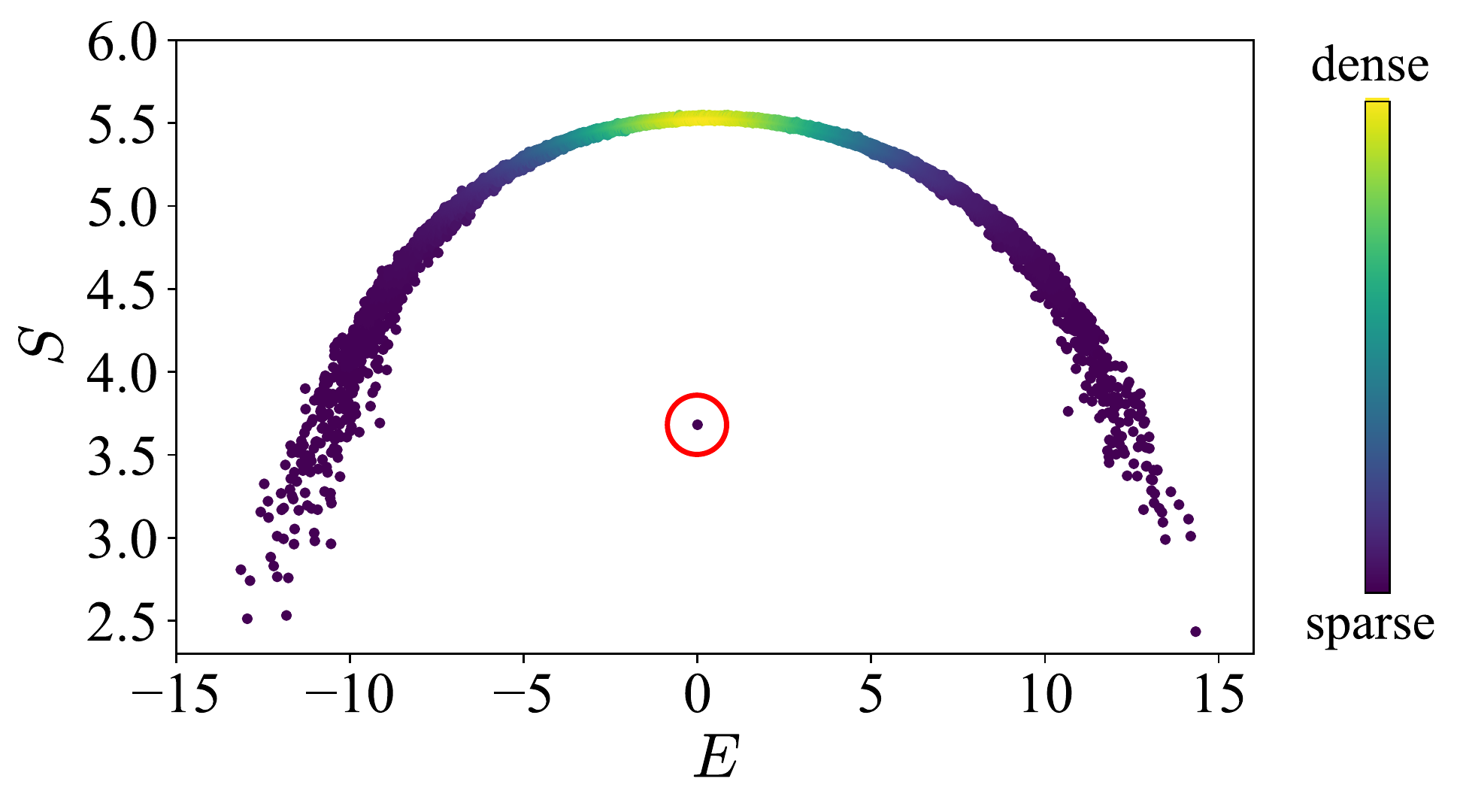}
        \end{minipage} \\
        \begin{minipage}[h]{\linewidth}
            \centering
            \subcaption{\label{fig:energy_vs_corrfunc}}
            \includegraphics[width=\linewidth]{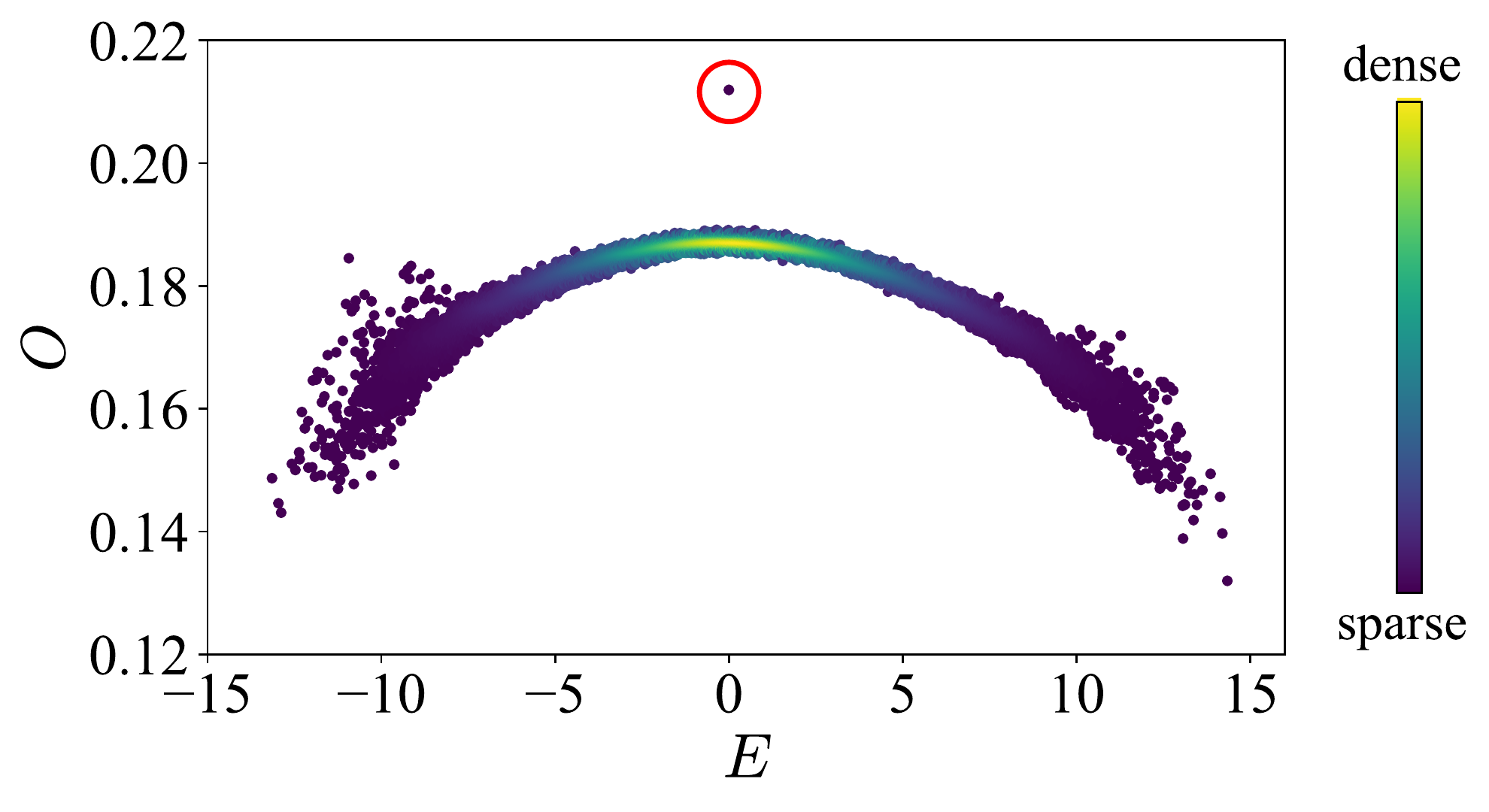} 
        \end{minipage}
    \end{tabular}
    \caption{
    \subref{fig:energy_vs_ee} Entanglement entropies in all eigenstates. The setup is the same as in Fig.~\ref{fig:level_spacing}.
    \subref{fig:energy_vs_corrfunc} The average of the correlation functions over all the bonds defined by Eq.~\eqref{eq:correlation_function} for each eigenstate.
    In each panel, the point enclosed by a red circle indicates the position of a QMBS.
    } \label{fig:eth_violation}
\end{figure}
\par
A violation of the ETH can be observed in the scaling of the entanglement entropy.
For a partition of the lattice into two subsystems $A$ and $B$, the entanglement entropy of a state $\ket{\psi}$ is defined as $S = - \Tr_A\left(\rho_A \ln{\rho_A}\right)$, where $\rho_A = \Tr_B\left(\dyad{\psi}\right)$ is the reduced density matrix of $A$. 
The symbols $\Tr_A$ and $\Tr_B$ denote traces over the subsystems $A$ and $B$, respectively.
It is known that the entanglement entropy of a thermal eigenstate obeys a volume law~\cite{mori2018thermalization}.
In contrast, if an eigenstate has a lower entanglement entropy than other eigenstates obeying a volume law even if its energy is far from the spectral edges, 
it signals that the state is a nonthermal QMBS.
Figure~\ref{fig:energy_vs_ee} shows the numerical result obtained by exact diagonalization with the same setup as in Fig.~\ref{fig:level_spacing}.
Here, the subsystems $A$ and $B$ are the left and right halves of the system, as shown in Fig.~\ref{fig:two-dim lattice}.
Clearly, there is a low-entanglement state isolated from the other, indicating the existence of a QMBS. 
The state has zero energy and is identified as $\ket{\Psi_5}$. 
Similar results are obtained for different particle-number sectors in which $\ket{\Psi_k}$ ($k = 1, 2, \dots$) appear as QMBS.
\par
One can also detect a violation of the ETH by observing expectation values of physical quantities.
We show the numerical result for the average value of the density-density correlation functions in Fig.~\ref{fig:energy_vs_corrfunc}.
The density-density correlation function is defined as 
\begin{align}
    O_{x, y} = \bra{\psi} n_x n_y \ket{\psi}
\end{align}
for a normalized state $\ket{\psi}$, and its average over all the bonds is given by
\begin{align}
    O = \frac{1}{|\mathcal{B}|} \sum_{\langle x, y\rangle \in \mathcal{B}} O_{x, y}. \label{eq:correlation_function}
\end{align}
The ETH states that, in an energy shell, the expectation value of any local observable in an energy eigenstate coincides with that obtained by the corresponding micro- canonical ensemble.
However, in Fig.~\ref{fig:energy_vs_corrfunc}, we see that there is an outlier at zero energy, which implies a violation of the ETH caused by QMBS.
\subsection{\label{subsec:dynamics} Dynamics}
To investigate further the nonthermal features of QMBS, we study the quench dynamics of the system. 
The state of the system at time $t$ is given by 
\begin{align}
    \ket{\Psi(t)} = \exp(-i H t) \ket{\Psi(0)},
\end{align}
where $\ket{\Psi(0)}$ is the initial state and $H$ is the Hamiltonian defined by Eq.~\eqref{eq:hamiltonian} with non-uniform matrix elements of $\mathsf{T}$ and $A_x$.
We discuss the dynamics of the fidelity and entanglement entropy starting from two kinds of initial states. 
The fidelity between the initial state $\ket{\Psi(0)}$ and the evolved state $\ket{\Psi(t)}$ is defined by \begin{align}
    F(t) = \left| \braket{\Psi(t)}{\Psi(0)}\right|. \label{eq:fidelity}
\end{align}
The dynamics of the entanglement entropy is as follows: 
we divide the lattice system into $A$ and $B$, and we define the reduced density matrix of $A$ at time $t$ as
\begin{align}
    \rho_A(t) = \Tr_B\left[\, \dyad{\Psi(t)} \,\right].
\end{align}
The entanglement entropy between $A$ and $B$ at time $t$ is then defined as 
\begin{align}
    S(t) &= - \Tr_A \left[ \, \rho_A(t) \ln \rho_A(t) \, \right]. \label{eq:dynamical_entanglement_entropy}
\end{align}
In our numerical simulation, we consider the two-dimensional lattice shown in Fig.~\ref{fig:two-dim lattice}.
We let the system evolve from two kinds of initial states for different particle numbers, and we calculate the dynamics for each initial state using exact diagonalization.
The first type of initial state is a product state defined by 
\begin{align}
    \ket{\Psi^{(\mathrm{prod})}_{N}} =
    c_N^\dag c_{N-1}^\dag \cdots c_1^\dag \ket{\mathrm{vac}}, \label{eq:product_state}
\end{align}
where $N$ is the number of particles, and the lattice sites are labeled as shown in Fig.~\ref{fig:two-dim lattice}. 
The second type takes the form \begin{align}
    \ket{\Psi^{(\mathrm{uni})}_{k}} = \frac{1}{\mathcal{N}}\left(\sum_{\langle x, y\rangle \in \mathcal{B}}c_x c_y\right)^k \ket{\overline{\mathrm{vac}}}, \label{eq:uniform_pairs_of_holes}
\end{align}
where $\mathcal{N}$ is the normalization constant and $k$ is an integer.
Note that the state~\eqref{eq:uniform_pairs_of_holes} is the QMBS when the matrix elements of $\mathsf{Q}$ are constant.
Therefore, although this state is not an eigenstate of $H$ with non-uniform $\mathsf{Q}$, we expect that it is close to the zero-energy eigenstate.
\par
\begin{figure}[tb]
    \centering
    \begin{tabular}{cc}
        \begin{minipage}[h]{0.5\linewidth}
            \centering
            \subcaption{\label{fig:overlap_dynamics_8}}
            \includegraphics[width=\linewidth]{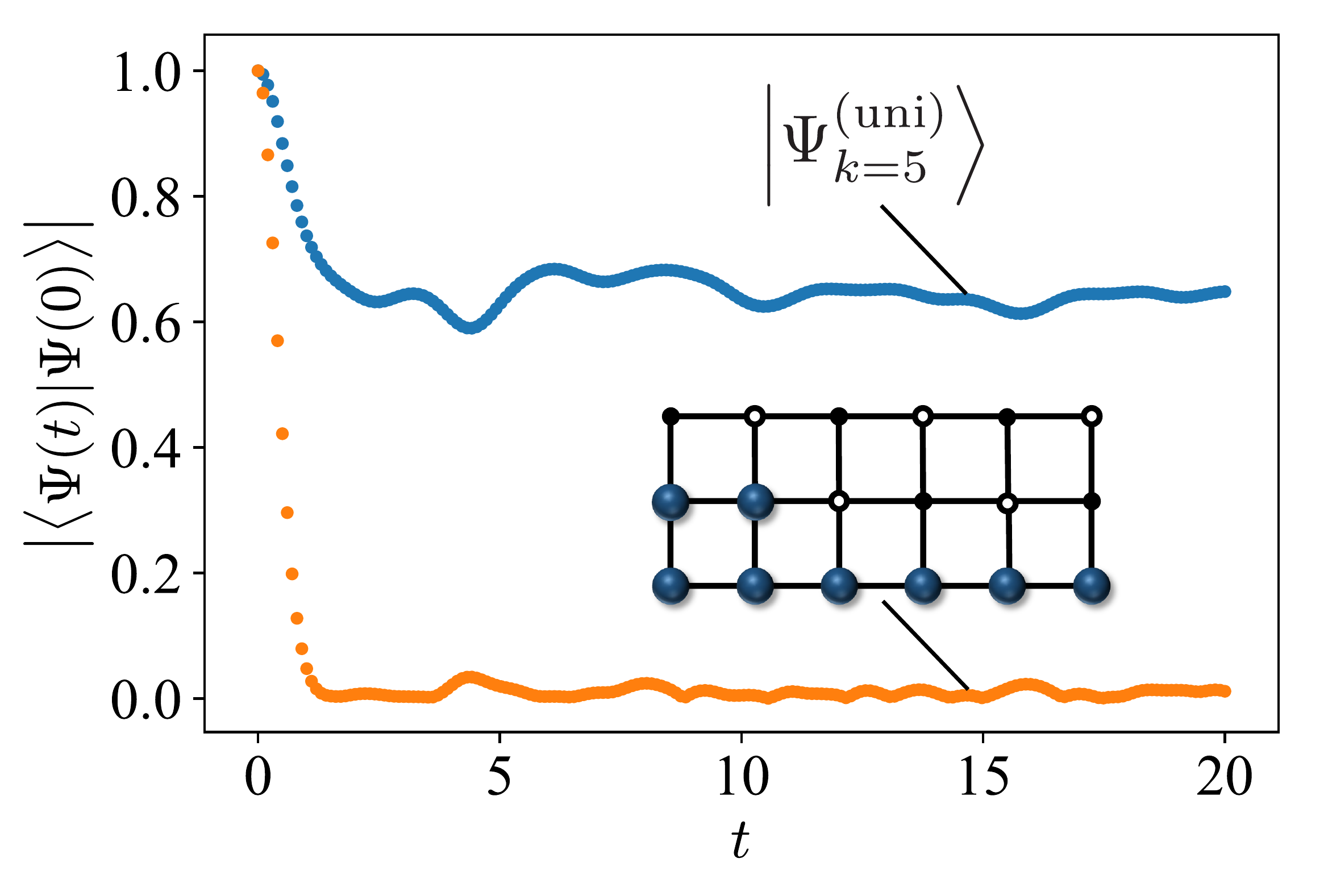} 
        \end{minipage} & 
        \begin{minipage}[h]{0.5\linewidth}
            \centering
            \subcaption{\label{fig:entanglement_entropy_dynamics_8}}
            \includegraphics[width=\linewidth]{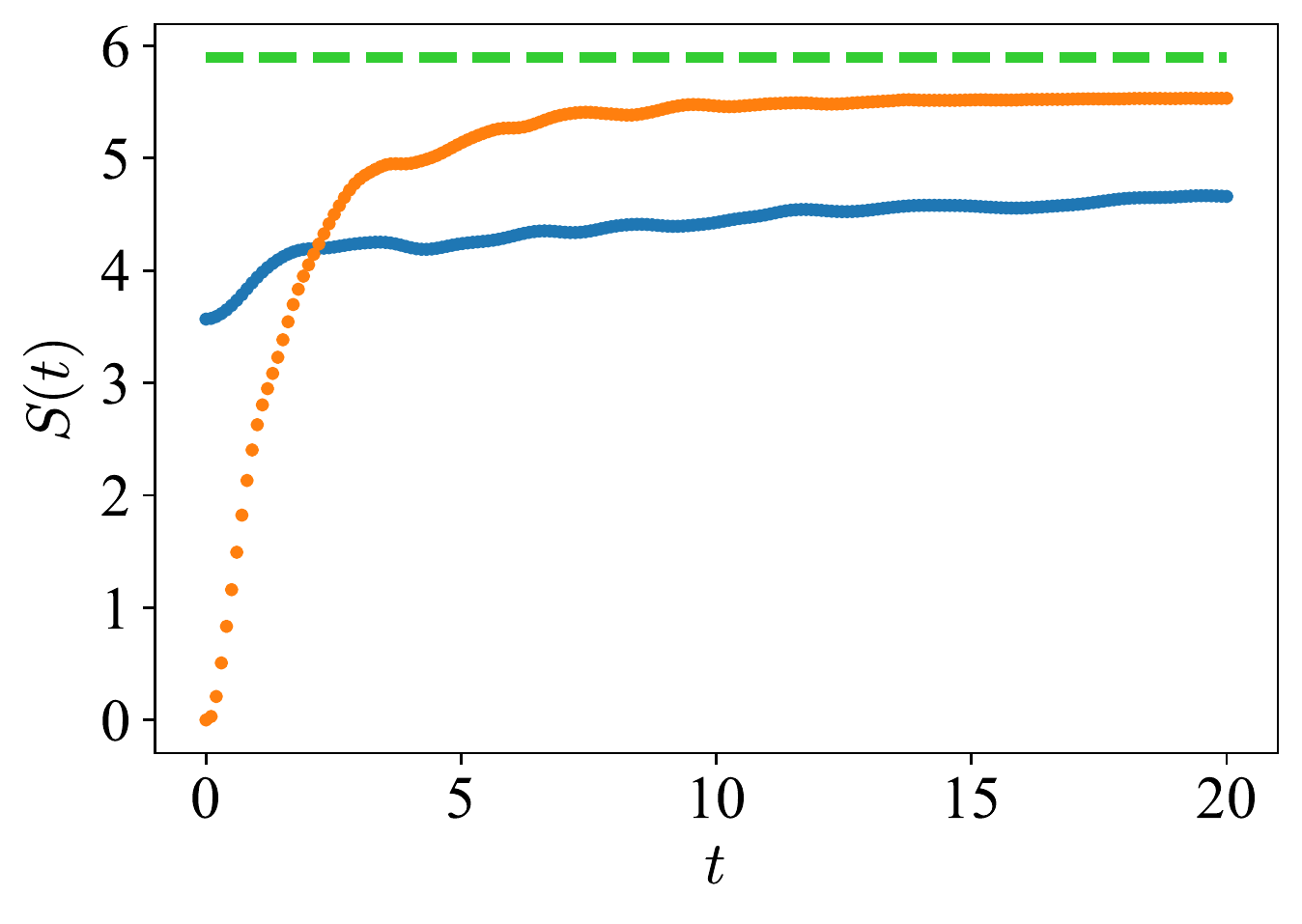} 
        \end{minipage} \\
        \begin{minipage}[h]{0.5\linewidth}
            \centering
            \subcaption{\label{fig:overlap_dynamics_14}}
            \includegraphics[width=\linewidth]{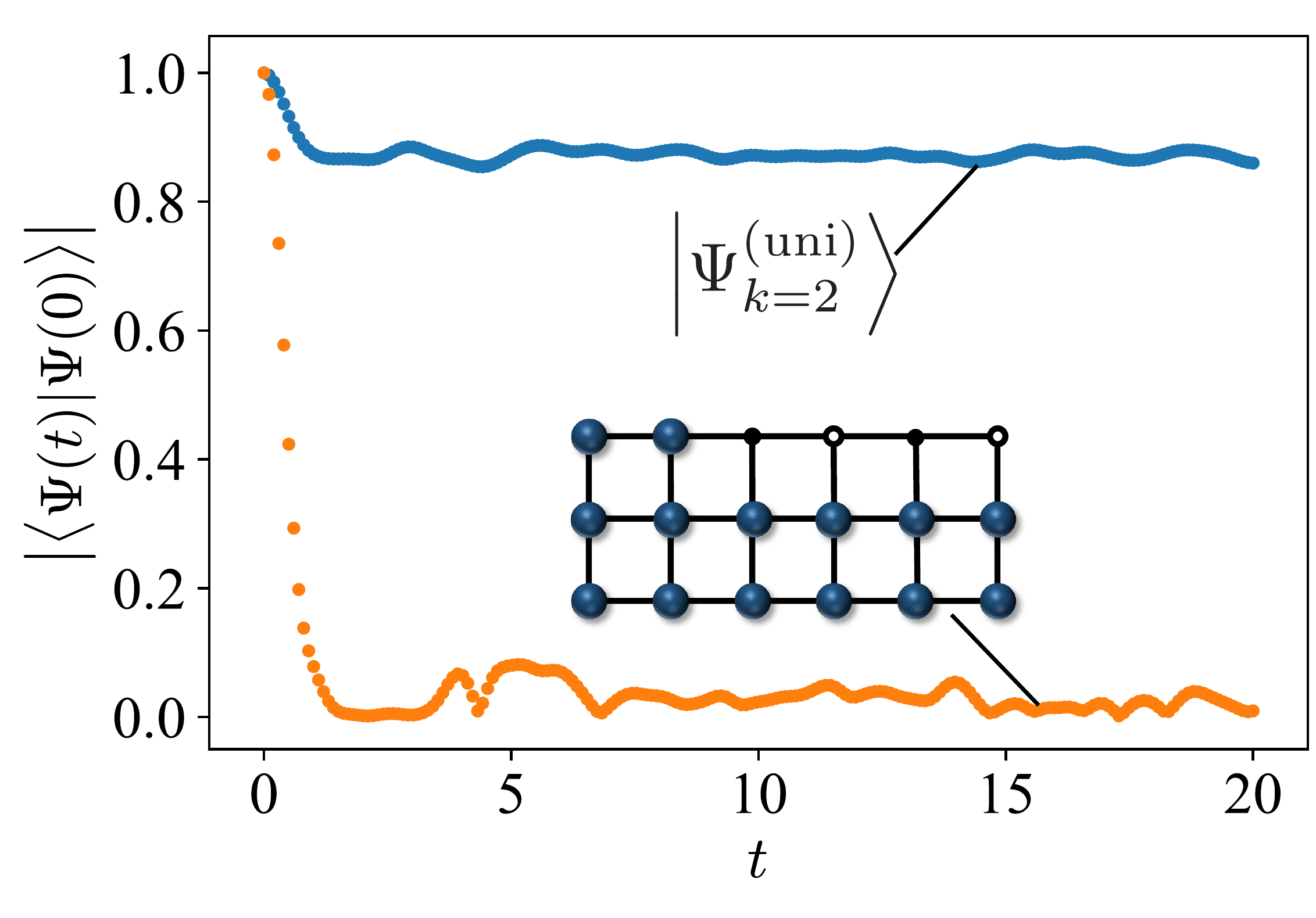} 
        \end{minipage} & 
        \begin{minipage}[h]{0.5\linewidth}
            \centering
            \subcaption{\label{fig:entanglement_entropy_dynamics_14}}
            \includegraphics[width=\linewidth]{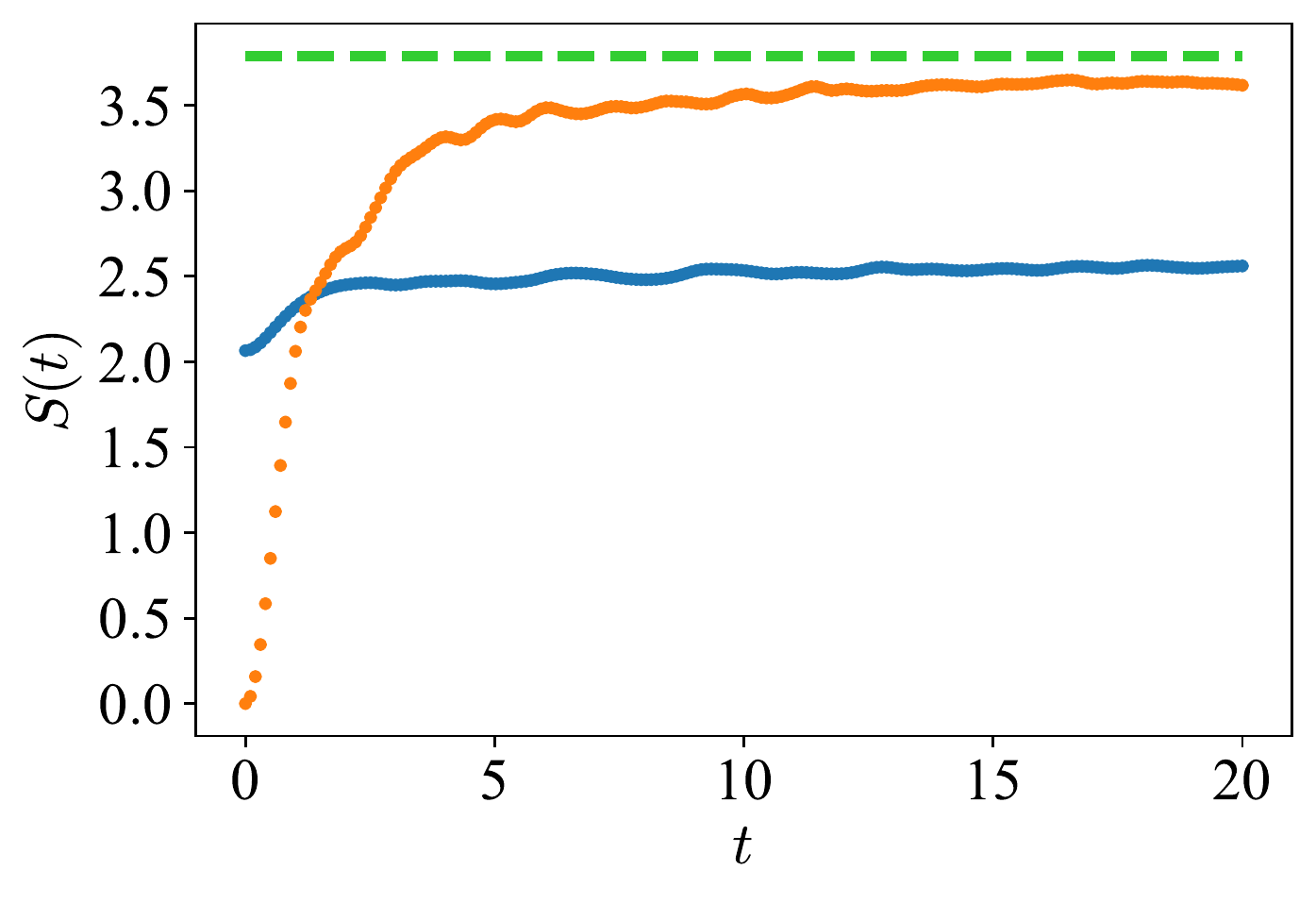} 
        \end{minipage}
    \end{tabular}
    \caption{ 
        Upper panel: Dynamics for $N=8$ with the same setup as in Fig.~\ref{fig:level_spacing}.
        The time dependence of \subref{fig:overlap_dynamics_8} fidelity and  \subref{fig:entanglement_entropy_dynamics_8} the entanglement entropy between the left and right halves.
        The orange (blue) line represents the dynamics starting from the state
       defined in Eq.~\eqref{eq:product_state} [\eqref{eq:uniform_pairs_of_holes}].
       The green dashed line indicates the average entanglement entropy of random states given by Eq.~\eqref{eq:average entropy}.
       Lower panel: Same as the upper panel for $N=14$.
        The time dependence of \subref{fig:overlap_dynamics_14} fidelity and \subref{fig:entanglement_entropy_dynamics_14} the entanglement entropy.
        The slight differences between the green dashed line and the orange lines at late times in \subref{fig:entanglement_entropy_dynamics_8} and \subref{fig:entanglement_entropy_dynamics_14} can be attributed to finite-size corrections to Eq.~\eqref{eq:average entropy}.
        } \label{fig:dynamics}
\end{figure}

Figures~\ref{fig:overlap_dynamics_8} and \ref{fig:overlap_dynamics_14} show the dynamics of fidelity for the same two-dimensional model as in Fig.~\ref{fig:level_spacing} for $N = 8$ ($k=5$) and $N = 14$ ($k=2$).
For the product state, we can see that it decays rapidly to zero.
On the other hand, for $\ket{\Psi_{k}^{(\mathrm{uni})}}$, it remains nonzero at late times, suggesting that the state $\ket{\Psi_{k}^{(\mathrm{uni})}}$ does not immediately thermalize due to significant overlap with the QMBS given in Eq.~\eqref{eq:qmbs}.
\par
Figures~\ref{fig:entanglement_entropy_dynamics_8} and \ref{fig:entanglement_entropy_dynamics_14} show the time dependence of the bipartite entanglement \eqref{eq:dynamical_entanglement_entropy} for $N = 8$ ($k=5$) and $N = 14$ ($k=2$).
Here, $A$ and $B$ are again the left and right halves of the lattice, respectively. 
For the product state, the entanglement entropy grows rapidly and saturates near the average entanglement entropy of random states with fixed particle number $N$~\cite{vidmar2017entanglement,bianchi2019typical,bianchi2022volume} 
\begin{align}
    S_{N} 
    &= \frac{1}{2}\left[(n-1) \ln (1-n) - n \ln n \right] |\Lambda| \nonumber \\
    & - \sqrt{\frac{n (1-n)}{2\pi}} \left| \ln \left(\frac{1-n}{n}\right)\right| \sqrt{|\Lambda|} + \frac{1 - 2 \ln 2}{4}, \label{eq:average entropy}
\end{align}
where $n = N/|\Lambda|$.
However, for the state $\ket{\Psi_{k}^{(\mathrm{uni})}}$, the growth of the entanglement entropy is suppressed, indicating nonthermalizing dynamics.
The results for both the fidelity and the entanglement entropy provide strong evidence that the system exhibits non-ergodic properties due to the QMBS. 

\section{\label{sec:The parent Hamiltonian} Parent Hamiltonian of the QMBS}
So far, we have constructed the exact QMBS and looked at the nonthermal properties.
In this section, we give a parent Hamiltonian whose ground states are the QMBS, and we prove that there are no other ground states. 
\par
The parent Hamiltonian we consider is given by
\begin{align}
    H_{\mathrm{par}} &= \sum_{x \in \Lambda} B_x h_x, \label{eq:parent Hamiltonian}\\
    h_x &= \left(\sum_{y \in \Lambda} q_{x, y} c_y^\dag\right)  c_x c_x^\dag \left(\sum_{y' \in \Lambda} q_{x, y'} c_{y'}\right), 
\end{align}
where all the $B_x$ are positive.
Since each $h_x$ is positive-semidefinite, the Hamiltonian $H_{\mathrm{par}}$ is also positive-semidefinite.
As in the scarred model, we assume that the lattice $(\Lambda, \mathcal{B})$ is bipartite, $\Lambda = \Lambda^{(1)} \cup \Lambda^{(2)}$.
We also assume that the matrix $\mathsf{Q} = \left(q_{x, y}\right)_{x, y \in \Lambda}$ is real skew-symmetric and that $q_{x, y}$ can be nonzero only when $\langle x, y\rangle \in \mathcal{B}$.
The parent Hamiltonian $H_{\rm par}$ is identical to Eq.~\eqref{eq:correlated_hopping_term} except that the coefficients $B_x$ are all positive.
Therefore, we can see that the states~\eqref{eq:qmbs} are zero-energy eigenstates of $H_{\mathrm{par}}$ in the same way as in Sec.~\ref{sec:qmbs}. 
Since $H_{\mathrm{par}}$ is positive-semidefinite, they are the ground states of $H_{\mathrm{par}}$. 
Under certain conditions, one can show that there are no other ground states. More precisely, we can prove the following theorem:

\smallskip
\textit{Theorem.} Assume that $\Lambda^{(1)}$ and $\Lambda^{(2)}$ have the same number of sites and $\mathsf{Q}$ is regular and connected, i.e., all $x \neq y \in \Lambda$ are connected via non-vanishing matrix elements of $\mathsf{Q}$.
Then the zero-energy ground states of $H_{\mathrm{par}}$ in the whole Fock space $\mathcal{V}$ are $(|\Lambda|/2 + 1)$-fold degenerate and written as $\ket{\Psi_k}$ defined in Eq.~\eqref{eq:qmbs} ($k = 0, 1, \dots, |\Lambda|/2$).
\smallskip

The lattice system in Fig.~\ref{fig:two-dim lattice} is an example that satisfies the assumption of the above theorem since the sublattices $\Lambda^{(1)}$ and $\Lambda^{(2)}$ have the same size. 
The theorem establishes that the QMBS can be prepared as the unique ground states of the parent Hamiltonian. 
We note that our model has much in common with the model studied in~\cite{tanaka2008ferromagnetic}, and the proof goes along the same lines as the proof of Proposition 2.1 in that paper.
The proof of the above theorem is given in Appendix~\ref{appendix:Proof of the uniqueness}.

\section{\label{sec:summary}Summary}
We have constructed and studied a class of spinless fermionic models with QMBS on a wide class of lattice systems, including higher-dimensional ones. 
We have also investigated the nonthermal properties of the QMBS through entanglement entropy and correlation functions. 
By examining the dynamics starting from different initial states, we confirmed that simple direct product states immediately thermalize, whereas the states with significant overlap with the QMBS exhibit a nonthermal behavior.
Furthermore, we have identified a parent Hamiltonian for which the QMBS we constructed are the unique ground states.

\begin{acknowledgments}
We are grateful to Leonardo Mazza and Akinori Tanaka for fruitful discussions. 
We used QuSpin~\cite{weinberg2017quspin,weinberg2019quspin} to calculate the entanglement entropy and physical quantities, as well as to simulate the quantum dynamics.
K.T. was supported by JSPS KAKENHI Grants No. 21J11575. 
H.K. was supported by MEXT KAKENHI Grant-in-Aid for Transformative Research Areas A ``Extreme Universe" No. JP21H05191, 
JSPS KAKENHI Grant No. JP18K03445, and the Inamori Foundation. 
\end{acknowledgments}

\appendix
\section{\label{appendix:particle-hole transformation}Particle-hole transformation}
We define the particle-hole transformation and derive another form of the Hamiltonian~\eqref{eq:hamiltonian}.
For each $x \in \Lambda$, let
\begin{align}
    U_x = c_x - c_x^\dag,
\end{align}
and we then define the unitary operator for the particle-hole transformation by
\begin{align}
    U = \prod_{x \in \Lambda} U_x.
\end{align}
The annihilation and creation operators transform as
\begin{align}
    U^\dag c_x U &= (-1)^{|\Lambda|} c_x^\dag, \\ 
    U^\dag c_x^\dag U &= (-1)^{|\Lambda|} c_x.
\end{align}
We see that the Hamiltonian \eqref{eq:hamiltonian} is transformed into another Hamiltonian as
\begin{align}
    \tilde{H} &= U^\dag H U = \tilde{H}_{\mathrm{hop}} + \tilde{H}_{\mathrm{cor}}, \label{eq:transformed hamiltonian}\\ 
    \tilde{H}_{\mathrm{hop}} &= U^\dag H_{\mathrm{hop}} U = - \sum_{x, y \in \Lambda} t_{x, y} c_x^\dag c_y, \\
    \tilde{H}_{\mathrm{cor}} &= U^\dag H_{\mathrm{cor}} U \nonumber \\
    &= - \sum_{x \in \Lambda} A_x \left(\sum_{y \in \Lambda} q_{x, y} c_y^\dag\right) c_x^\dag c_x \left(\sum_{y' \in \Lambda} q_{x, y'} c_{y'}\right) \nonumber \\
    & \ \ \ \ - \sum_{x \in \Lambda} A_x \left(\mathsf{Q}^2\right)_{x, x} c_x^\dag c_x.
\end{align}
The QMBS for the transformed Hamiltonian~\eqref{eq:transformed hamiltonian} are obtained as
\begin{align}
    \ket*{\tilde{\Psi}_k} = U^\dag \ket{\Psi_k}.
\end{align}
Clearly, they are zero-energy eigenstates of $\tilde{H}$. 
With the particle-hole transformation, the operator $Q$ changes to
\begin{align}
    \tilde{Q} = U^\dag Q U = \sum_{x, y \in \Lambda} q_{x, y} c_x^\dag c_y^\dag.
\end{align}
Thus we find that 
\begin{align}
    \ket*{\tilde{\Psi}_k} =  (-1)^{|\Lambda|} \tilde{Q}^k \ket{\mathrm{vac}},
\end{align}
where we have used
\begin{align}
    U^\dag \left(\prod_{x \in \Lambda} c_x^\dag\right) \ket{\mathrm{vac}} = (-1)^{|\Lambda|} \ket{\mathrm{vac}}.
\end{align}

\section{\label{appendix:generalization}
Generalization of our construction to general lattices}
\begin{figure}[tb]
\includegraphics[width=\columnwidth]{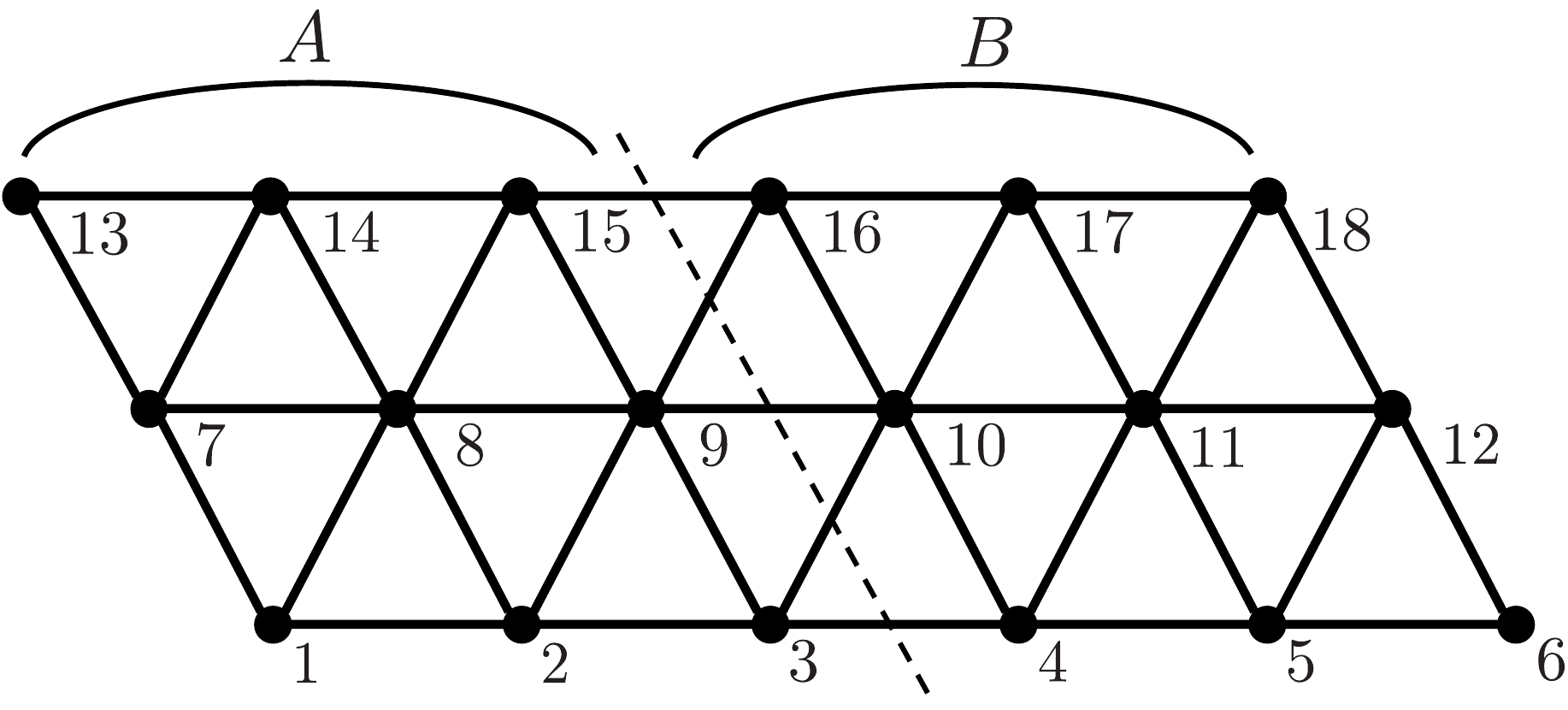}
\caption{An example of a non-bipartite lattice (triangular lattice).
Regions $A$ and $B$ are defined for the calculation of the entanglement entropy.
The integers are labels for the lattice sites.
} \label{fig:traiangular_lattice}
\end{figure}

We generalize our construction of models with QMBS to general lattices that are not necessarily bipartite (see, e.g., Fig.~\ref{fig:traiangular_lattice}).
Let $(\Lambda, \mathcal{B})$ be an arbitrary lattice. The Hamiltonian $H$ is given by 
\begin{align}
    H &= H_{\mathrm{hop}} + H_{\mathrm{cor}}, \label{eq:app_hamiltonian}\\ 
    H_{\mathrm{hop}} &= \sum_{x, y \in \Lambda} t_{x, y} c_x^\dag c_y, \\
    H_{\mathrm{cor}} &= \sum_{x \in \Lambda} A_x \left(\sum_{y \in \Lambda} q_{x, y} c_y^\dag \right)c_x c_x^\dag \left(\sum_{y' \in \Lambda} q_{x, y'} c_{y'}\right).
\end{align}
Here, the hopping matrix $\mathsf{T}$ is assumed to be of the form 
\begin{align}
    \mathsf{T} = i \mathsf{K},
\end{align}
where $\mathsf{K}$ is a real skew-symmetric matrix, and the matrix element $\left(\mathsf{K}\right)_{x, y}$ can be nonzero only when $\langle x, y \rangle \in \mathcal{B}$.
The matrix $\mathsf{Q} = \left(q_{x,y}\right)_{x, y \in \Lambda}$ is defined as 
\begin{align}
    \mathsf{Q} = \mathsf{K}^n,
\end{align}
where $n$ is a positive odd integer.
\par
In a similar way as in Sec.~\ref{subsec:qmbs eigenstate}, the QMBS states are constructed as 
\begin{align}
    \ket{\Psi_k} = Q^k \ket{\overline{\mathrm{vac}}}, \label{eq:app_qmbs_states}
\end{align}
where
\begin{align}
    Q = \sum_{x, y \in \Lambda} q_{x, y} c_x c_y.
\end{align}
Repeating calculations similar to those in Sec.~\ref{sec:qmbs}, we obtain the following commutation relations,  
\begin{align}
    [H_{\mathrm{hop}}, Q] &= 0, \\ 
    \left[c_x^\dag \left(\sum_{y \in \Lambda} q_{x, y} c_y\right), Q\right] &= 0 \ \ \text{for all} \ x \in \Lambda,
\end{align}
and $H_{\mathrm{hop}} \ket{\overline{\mathrm{vac}}} = H_{\mathrm{cor}} \ket{\overline{\mathrm{vac}}} = 0$.
Thus we see that the states of the form Eq.~\eqref{eq:app_qmbs_states} are eigenstates of the Hamiltonian~\eqref{eq:app_hamiltonian} with zero energy.
\par
\begin{figure}[tb]
\centering
\includegraphics[width=0.35\textwidth]{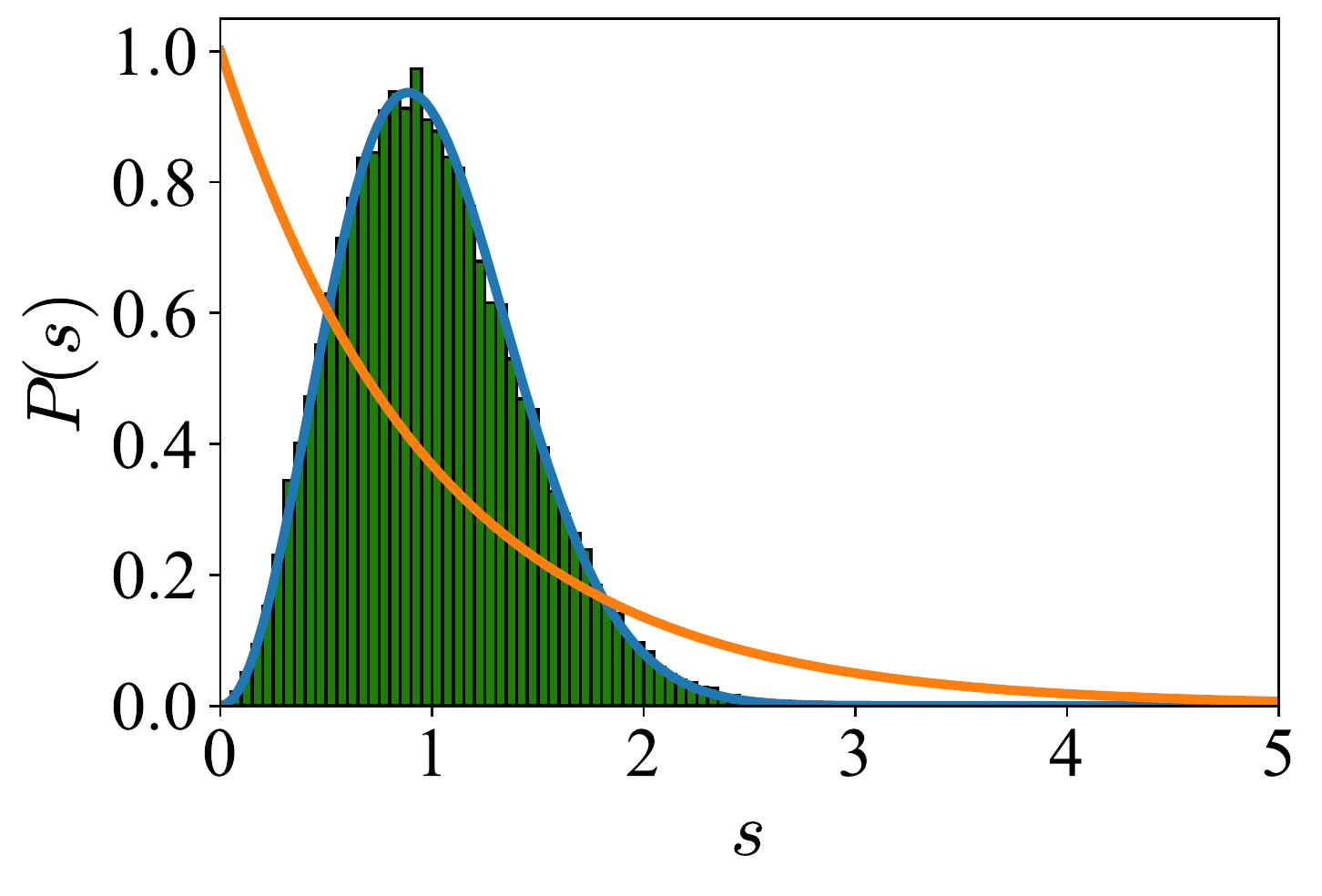}
\caption{The level statistics in the middle half of the spectrum of the Hamiltonian in Eq.~\eqref{eq:app_hamiltonian} on the triangular lattice shown in Fig.~\ref{fig:traiangular_lattice}.
The number of lattice sites is 18, and the particle number $N$ is fixed to 8.
The matrix $\mathsf{Q}$ is given by $\mathsf{Q} = \mathsf{K}$, and the matrix elements of $\mathsf{K}$ and $A_x$ are chosen uniformly at random from $[0.5, 1.5]$ and $[-0.5, 0.5]$, respectively.
The blue (orange) line is the GUE Wigner-Dyson distribution (the Poisson distribution). 
} \label{fig:level_spacing_triangular_lattice}
\end{figure}

To check the nonintegrability, we perform the energy level statistics.
For nonintegrable systems without time-reversal symmetry, the level-spacing distribution is given by the GUE Wigner-Dyson distribution $P(s) = (32/\pi^2) s^2 e^{4 s^2/\pi}$, whereas for integrable systems it is given by the Poisson distribution $P(s)= e^{-s}$.
We show the distribution of the level spacings of the model on a triangular lattice in Fig.~\ref{fig:level_spacing_triangular_lattice}.
The histogram confirms that the distribution is well described by the GUE Wigner-Dyson distribution.
We also compute the level-spacing ration $r_i$, and it is known that the average value of $r_i$ is $\langle r \rangle \simeq 0.60266$ for GUE and $\langle r \rangle \simeq 0.38629$ for the Poisson distribution~\cite{atas2013distribution}.
The average value of $r_i$ calculated from the histogram in Fig,~\ref{fig:level_spacing_triangular_lattice} is $0.59998...$, which is close to that of the GUE Wigner-Dyson distribution.
Therefore, we can conclude that our model is nonintegrable.
\par
\begin{figure}[tb]
    \centering
    \begin{tabular}{c}
        \begin{minipage}[h]{\linewidth}
            \centering
            \subcaption{\label{fig:energy_vs_ee_triangular_lattice}}
            \includegraphics[width=\linewidth]{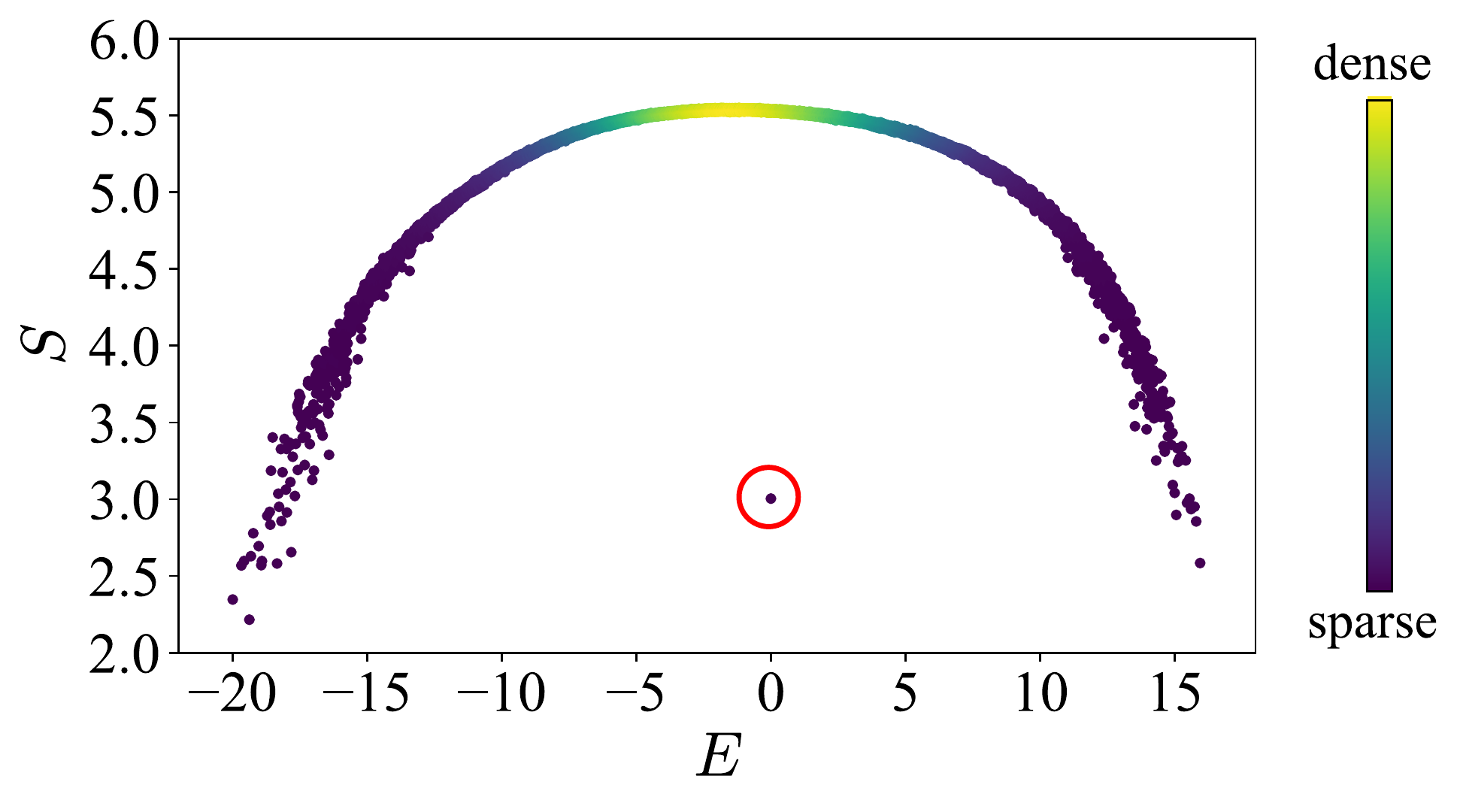} 
        \end{minipage} \\
        \begin{minipage}[h]{\linewidth}
            \centering
            \subcaption{\label{fig:energy_vs_corr_func_triangular_lattice}}
            \includegraphics[width=\linewidth]{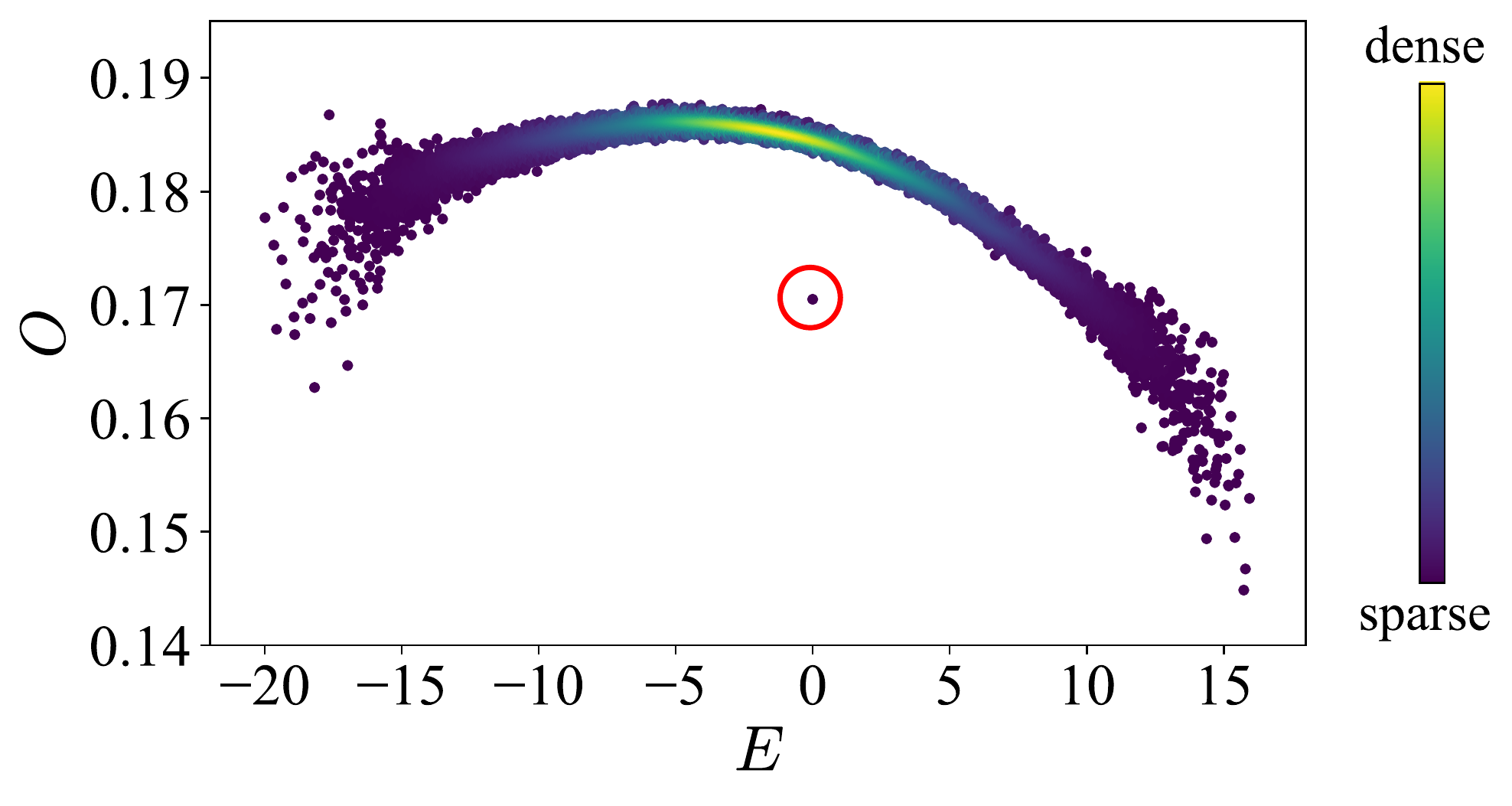} 
        \end{minipage}
    \end{tabular}
    \caption{\subref{fig:energy_vs_ee_triangular_lattice} Entanglement entropies in all eigenstates. The setup is the same as in Fig.~\ref{fig:level_spacing_triangular_lattice}.
    \subref{fig:energy_vs_corr_func_triangular_lattice} 
    The average of the correlation functions over all the bonds defined by Eq.~\eqref{eq:correlation_function} for each eigenstate. 
    In each panel, the point enclosed by a red circle indicates the position of a QMBS.} \label{fig:eth_violation_triangular_lattice}
\end{figure}

We also study the entanglement entropy and the average of the density-density correlation functions to see if the QMBS violate the ETH.
Figure~\ref{fig:eth_violation_triangular_lattice} shows the numerical results for the same setting as in Fig.~\ref{fig:level_spacing_triangular_lattice}. 
Figure~\ref{fig:energy_vs_ee_triangular_lattice} shows the entanglement entropies of the eigenstates.
To calculate the entanglement entropy, we divide the lattice system into two systems $A$ and $B$, as shown in Fig.~\ref{fig:traiangular_lattice}.
We can see that there is an eigenstate having a lower entanglement entropy than the others in the middle of the spectrum.
This state coincides with $\ket{\Psi_{k}}$ with $k=5$.
Figure~\ref{fig:energy_vs_corr_func_triangular_lattice} shows the average values of the correlation functions defined in Eq.~\eqref{eq:correlation_function} for each eigenstate, where there is an isolated point at zero energy.
These results suggest that the zero-energy eigenstates~\eqref{eq:app_qmbs_states} are nonthermal.
\par
\begin{figure}[tb]
    \centering
    \begin{tabular}{cc}
        \begin{minipage}[h]{0.5\linewidth}
            \centering
            \subcaption{\label{fig:overlap_dynamics_triangular_lattice_8}}
            \includegraphics[width=\linewidth]{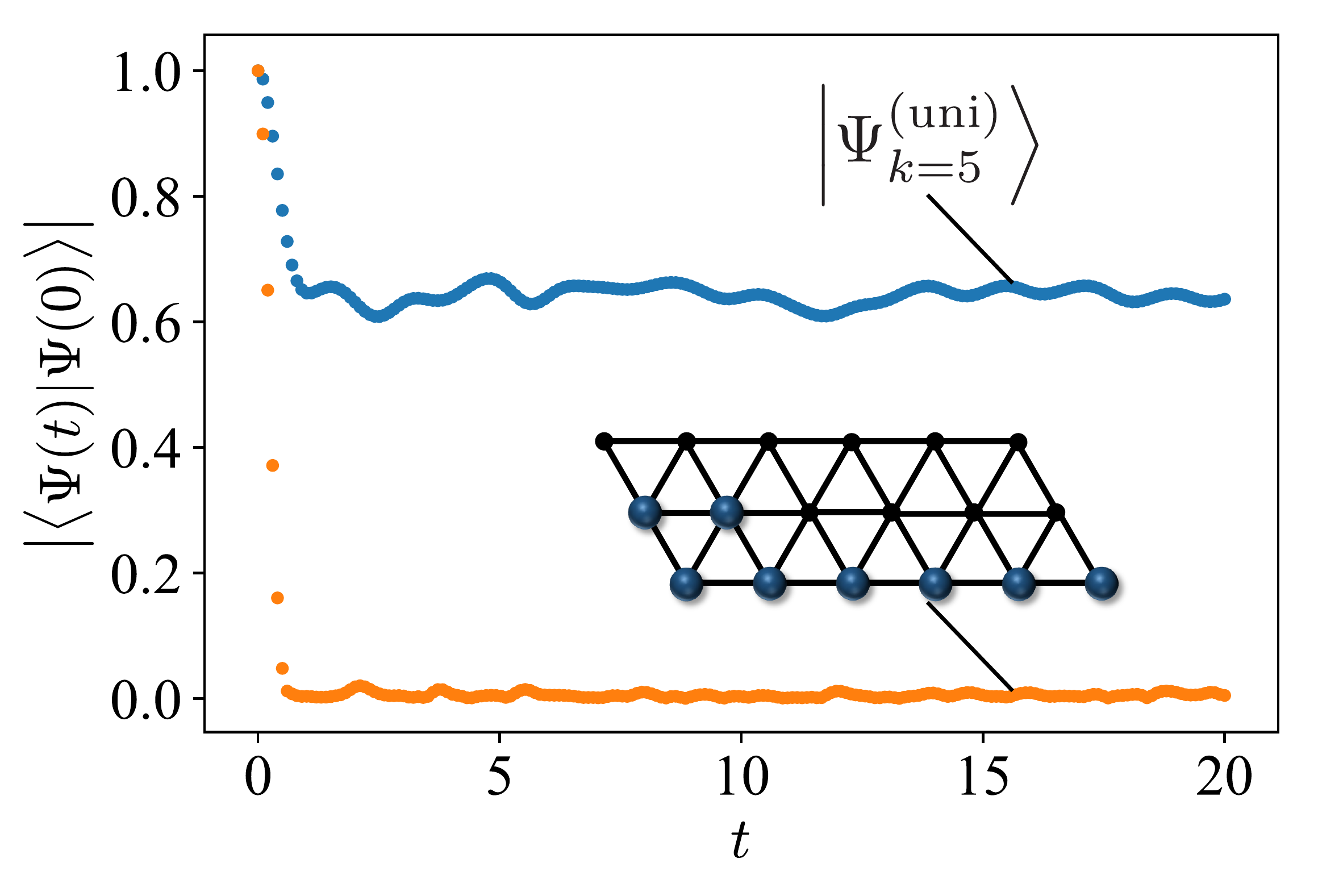} 
        \end{minipage} & 
        \begin{minipage}[h]{0.5\linewidth}
            \centering
            \subcaption{\label{fig:entanglement_entropy_dynamics_triangular_lattice_8}}
            \includegraphics[width=\linewidth]{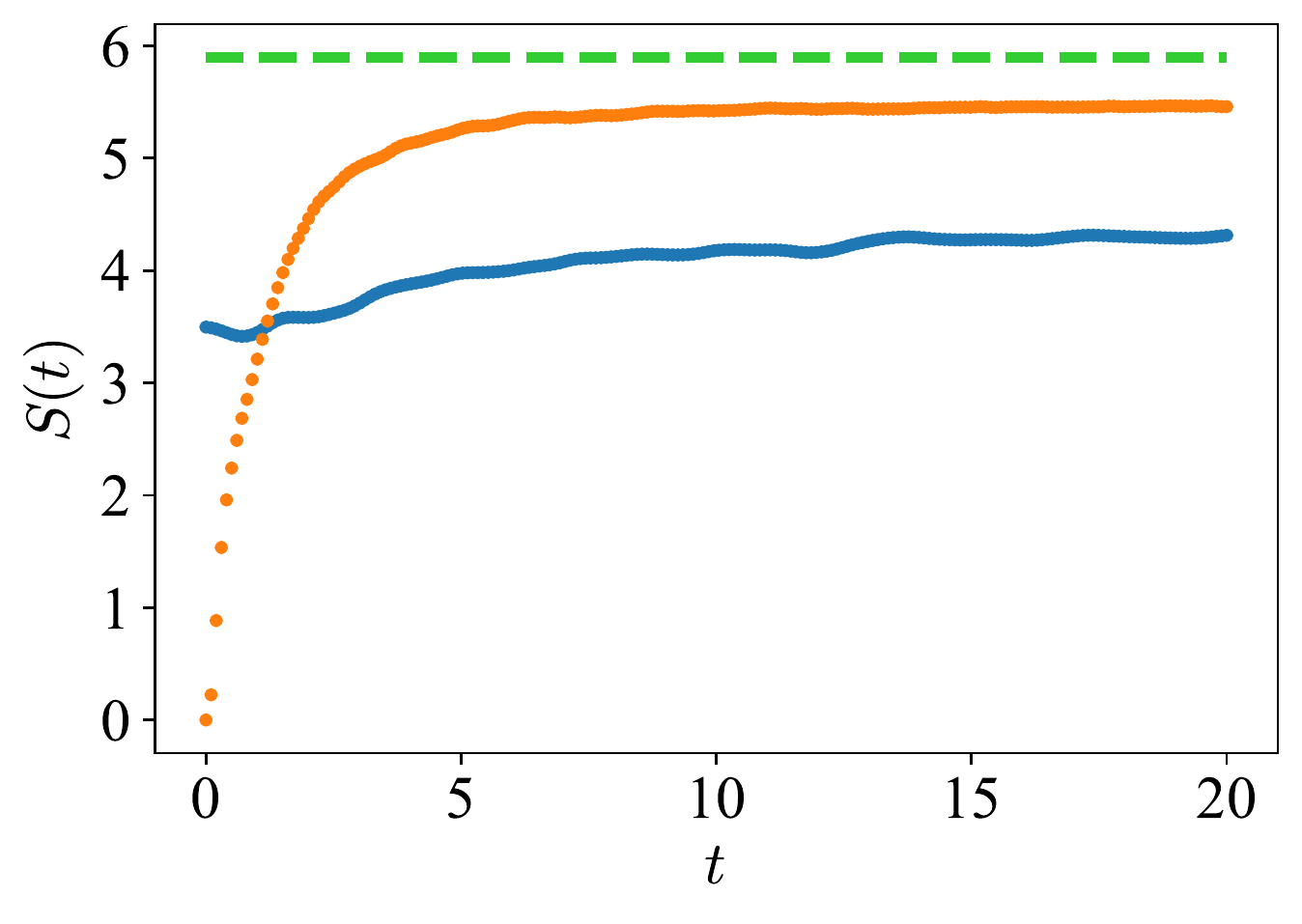} 
        \end{minipage} \\
        \begin{minipage}[h]{0.5\linewidth}
            \centering
            \subcaption{\label{fig:overlap_dynamics_triangular_lattice_14}}
            \includegraphics[width=\linewidth]{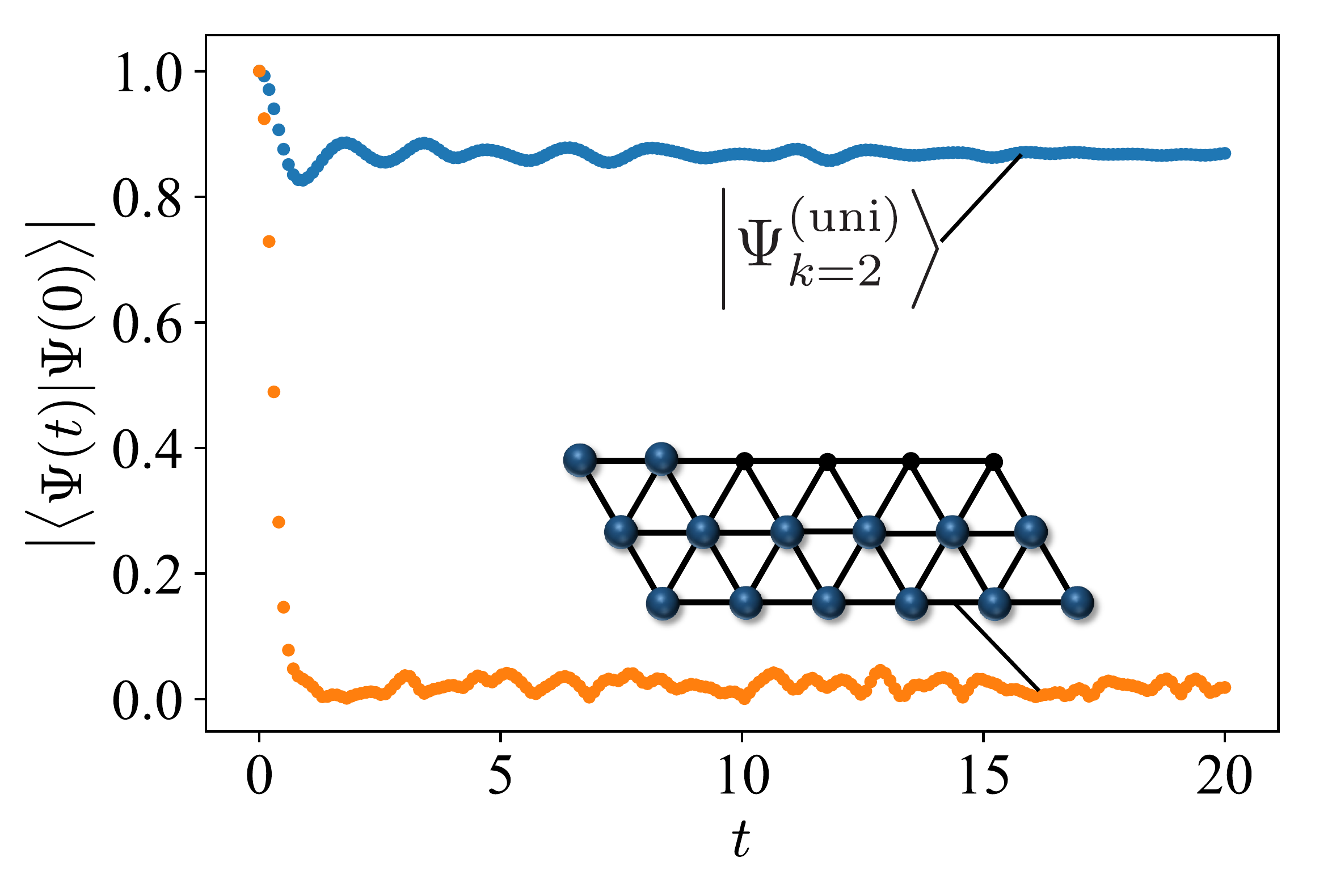} 
        \end{minipage} & 
        \begin{minipage}[h]{0.5\linewidth}
            \centering
            \subcaption{\label{fig:entanglement_entropy_dynamics_triangular_lattice_14}}
            \includegraphics[width=\linewidth]{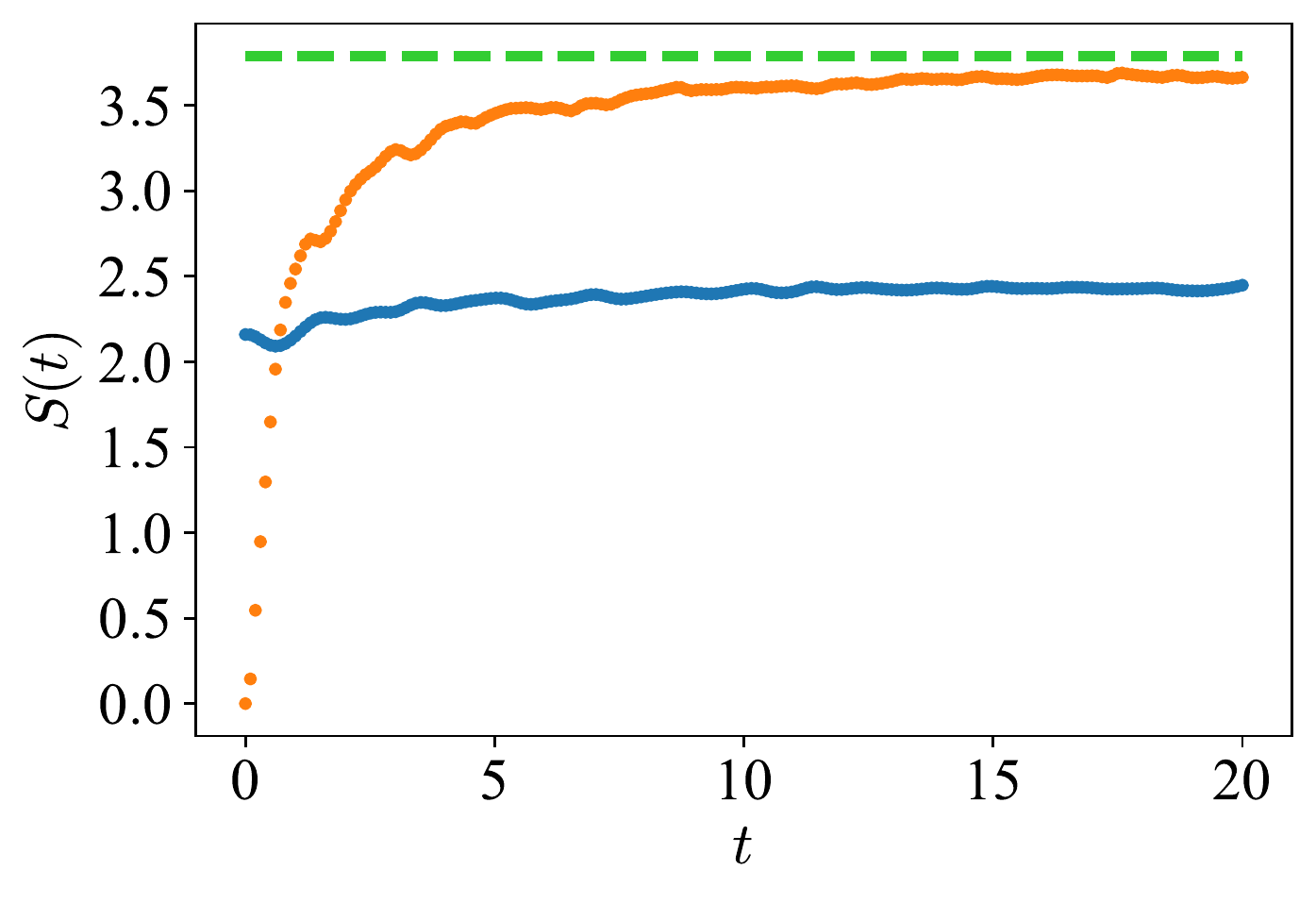} 
        \end{minipage}
    \end{tabular}
    \caption{\label{fig:dynamics_triangular_lattice} Upper panel: Dynamics for $N=8$ with the same setup as in Fig.~\ref{fig:level_spacing_triangular_lattice}.
    The time dependence of \subref{fig:overlap_dynamics_triangular_lattice_8} fidelity and \subref{fig:entanglement_entropy_dynamics_triangular_lattice_8} the entanglement entropy between the left and right halves.
    The orange (blue) line represents the dynamics starting from the state defined in Eq.~\eqref{eq:product_state}[\eqref{eq:uniform_pairs_of_holes}].
    The green dashed line indicates the average entanglement entropy of random states given by Eq.~\eqref{eq:average entropy}.
    Lower panel: Same as the upper panel for $N=14$.
    The time dependence of \subref{fig:overlap_dynamics_triangular_lattice_14} fidelity and \subref{fig:entanglement_entropy_dynamics_triangular_lattice_14} the entanglement entropy.
    The slight differences between the green dashed line and the orange lines at late times in \subref{fig:entanglement_entropy_dynamics_triangular_lattice_8} and \subref{fig:entanglement_entropy_dynamics_triangular_lattice_14} can be attributed to finite-size corrections to Eq.~\eqref{eq:average entropy}.
    }
\end{figure}

Finally, we examine the dynamics of the fidelity and the entanglement entropy, which are defined as in the previous case by Eqs.~\eqref{eq:fidelity} and \eqref{eq:dynamical_entanglement_entropy}.
We compare two kinds of initial states defined in the same form as Eqs.~\eqref{eq:product_state} and \eqref{eq:uniform_pairs_of_holes}.
The product state $\ket{\Psi_{N}^{(\mathrm{prod})}}$ is defined according to the order of the labels in Fig.~\ref{fig:traiangular_lattice}.
Figure~\ref{fig:dynamics_triangular_lattice} shows the dynamics of the fidelity and the entanglement entropy when $N=8$ and $14$ in the same setup as in Fig.~\ref{fig:level_spacing_triangular_lattice}.
The fidelity dynamics is shown in Figs.~\ref{fig:overlap_dynamics_triangular_lattice_8} and \ref{fig:overlap_dynamics_triangular_lattice_14}, where we can see that it rapidly relaxes to zero when starting from the product state, while it remains non-zero at late times when starting from the state Eq.~\eqref{eq:uniform_pairs_of_holes}.
Figures~\ref{fig:entanglement_entropy_dynamics_8} and \ref{fig:entanglement_entropy_dynamics_14} show the dynamics of the entanglement entropy.
It rapidly grows and saturates near the value of Eq.~\eqref{eq:average entropy} when starting from the product state, while it grows very slowly when starting from the state $\ket{\Psi_k^{(\mathrm{uni})}}$.
As in the previous case, these behaviors imply nonthermalizing dynamics of $\ket{\Psi_{k}^{(\mathrm{uni})}}$.
\section{\label{appendix:Q operator and pairing states} Pairing operator with zero energy}
We show that the operator $Q$ defined in Eq.~\eqref{eq:q operator} is a sum of pair operators, each of which commutes with $H_{\mathrm{hop}}$ in Eq.~\eqref{eq:hopping_term}. 
Since the matrix element $t_{x, y}$ is zero when $x$ and $y$ are in the same sublattice $\Lambda^{(1)}$ or $\Lambda^{(2)}$, the matrix $\mathsf{T}$ can be written in the form
\begin{align}
    \mathsf{T} = \begin{pmatrix}
     \mathsf{0} & \mathsf{M}_{1,2} \\
    \mathsf{M}_{1, 2}^{\mathrm{T}} &  \mathsf{0}
    \end{pmatrix},
\end{align}
where $\mathsf{M}_{1, 2}$ is a $|\Lambda^{(1)}| \times |\Lambda^{(2)}|$ matrix, and $\mathsf{M}_{1, 2}^{\mathrm{T}}$ denotes the transpose of $\mathsf{M}_{1, 2}$.
Now let $\mathsf{M}_{1, 2} = \mathsf{U} \mathsf{S} \mathsf{V}^{\mathrm{T}}$ be the singular value decomposition of $\mathsf{T}$, where $\mathsf{U}$ and $\mathsf{V}$ are $|\Lambda^{(1)}| \times |\Lambda^{(1)}|$ and $|\Lambda^{(2)}| \times |\Lambda^{(2)}|$ orthogonal matrices, respectively.
The matrix $\mathsf{S}$ is a $|\Lambda^{(1)}| \times |\Lambda^{(2)}|$ rectangular diagonal matrix with nonnegative diagonal elements.
In the following, we denote the number of nonzero diagonal elements of $\mathsf{S}$ by $r$, and let them be $\varepsilon_k > 0$ ($k=1, 2, \dots, r$).
With the decomposition, we have 
\begin{align}
    \mathsf{T} = \begin{pmatrix}
        \mathsf{U} & \mathsf{0}\\
         \mathsf{0} &  \mathsf{V}
    \end{pmatrix}
    \begin{pmatrix}
        \mathsf{0} & \mathsf{S} \\ 
        \mathsf{S}^{\mathrm{T}} & \mathsf{0}
    \end{pmatrix}
    \begin{pmatrix}
        \mathsf{U}^{\mathrm{T}} & \mathsf{0} \\
        \mathsf{0} & \mathsf{V}^{\mathrm{T}}
    \end{pmatrix}.
\end{align}
We note that there exists an orthogonal matrix $\mathsf{P}$ such that
\begin{align}
    \begin{pmatrix}
        \mathsf{0} & \mathsf{S} \\ 
        \mathsf{S}^{\mathrm{T}} & \mathsf{0}
    \end{pmatrix} = \mathsf{P}
    \begin{pmatrix}
        \begin{matrix}
            0 & \varepsilon_1 \\ 
            \varepsilon_1 & 0
        \end{matrix} & & & \\ 
        & \ddots & & \\
        & & \begin{matrix}
            0 & \varepsilon_r \\ 
            \varepsilon_r & 0
        \end{matrix} &  \\ 
         & & & \begin{matrix}
             0 & &  \\
             & \ddots & \\ 
             & & 0
         \end{matrix}
    \end{pmatrix}
    \mathsf{P}^{\mathrm{T}},
\end{align}
where matrix elements which are zero are left empty. 
Letting 
\begin{align}
    \mathsf{R} = \begin{pmatrix}
        \mathsf{U} & \mathsf{0}\\
        \mathsf{0} &  \mathsf{V}
    \end{pmatrix}
    \mathsf{P}
\end{align}
and $a_i = \sum_{x \in \Lambda} c_x (\mathsf{R})_{x, i}$, we find that $H_{\mathrm{hop}}$ can be expressed as 
\begin{align}
    H_{\mathrm{hop}} = \sum_{k=1}^r \varepsilon_k \left(a_{2k-1}^\dag a_{2k} + a_{2k}^\dag a_{2k-1}\right).
\end{align}
In the same way, the matrix $\mathsf{Q}$ can be represented as 
\begin{align}
    \mathsf{Q} = \mathsf{R}
    \begin{pmatrix}
        \begin{matrix}
            0 & \varepsilon_1 \\ 
            -\varepsilon_1 & 0
        \end{matrix} & & & & \\ 
        & \ddots & & \\
        & & \begin{matrix}
            0 & \varepsilon_r \\ 
            -\varepsilon_r & 0
        \end{matrix} &  \\ 
         & & & \begin{matrix}
             0 & & \\
             & \ddots & \\ 
             & & 0
         \end{matrix}
    \end{pmatrix}
    \mathsf{R}^{\mathrm{T}}.
\end{align}
Thus we see that the operator $Q$ can be rewritten as 
\begin{align}
    Q = 2 \sum_{k=1}^r a_{2k-1} a_{2k}.
\end{align}
To diagonalize $H_{\mathrm{hop}}$, we introduce 
\begin{align}
    b_{2k-1} &= \frac{a_{2k-1} + a_{2k}}{\sqrt{2}}, \\ 
    b_{2k} &= \frac{a_{2k-1} - a_{2k}}{\sqrt{2}},
\end{align}
for $k = 1, 2, \dots, r$. 
They satisfy 
\begin{align}
    \{b_i, b_j\} &= \{b_i^\dag, b_j^\dag\} = 0, \\
    \{b_i, b_j^\dag\} &= \delta_{i,j}
\end{align}
for $i, j = 1, 2, .\dots, 2r$. 
In terms of the new fermionic operators, $H_{\mathrm{hop}}$ takes the form
\begin{align}
    H_{\mathrm{hop}} = \sum_{k=1}^r \varepsilon_k(b_{2k-1}^\dag b_{2k-1} - b_{2k}^\dag b_{2k} ),
\end{align}
which shows that nonzero eigenvalues of $\mathsf{T}$ come in pairs, $\pm \varepsilon_k$.
In a similar manner, we see that 
\begin{align}
    Q = 2 \sum_{k=1}^r \varepsilon_k b_{2k} b_{2k-1}.
\end{align}
Because $[H_{\mathrm{hop}}, b_{2k-1}] = - \varepsilon_k b_{2k-1}$ and $[H_{\mathrm{hop}}, b_{2k}] = \varepsilon_k b_{2k}$, the pair operator $b_{2k} b_{2k-1}$ satisfies $[H_{\mathrm{hop}}, b_{2k} b_{2k-1}] = 0$.
This means that $b_{2k} b_{2k-1}$ is an operator that annihilates a pair of eigenmodes of $H_{\mathrm{hop}}$ whose energies add up to zero. 
\section{\label{appendix:Proof of the uniqueness} Proof of the uniqueness of the ground states}
We give a complete proof of the theorem in Sec.~\ref{sec:The parent Hamiltonian}.
\par
\textit{Proof}--- 
We define a set of operators 
\begin{align}
    q_{x} = \sum_{y \in \Lambda} q_{x, y} c_y \ \  \text{for all $x \in \Lambda$}.
\end{align}
Since the matrix $\mathsf{Q}$ is assumed to be regular, we can define its dual operators as 
\begin{align}
    \tilde{q}_x^\dag = \sum_{y \in \Lambda} c_y^\dag \left(\mathsf{Q}^{-1} \right)_{y, x}, 
\end{align}
which satisfy the following anticommutation relations,
\begin{align}
    \{q_x, \tilde{q}_y^\dag \} = \delta_{x, y}. \label{eq:anticommutation relation of q}
\end{align}
Note that the matrix element $q_{x, y}$ is zero if both $x$ and $y$ are in $\Lambda^{(1)}$ or $\Lambda^{(2)}$, and in this case, $\left(\mathsf{Q}^{-1}\right)_{x, y}$ is also zero if both $x$ and $y$ are in $\Lambda^{(1)}$ or $\Lambda^{(2)}$.
\par
Let $\ket{\Psi_{\mathrm{GS}}}$ be a zero-energy ground state of $H_{\mathrm{par}}$ and assume that it is an $N$-particle state.
For arbitrary pairs of subsets $X, X' \subset \Lambda^{(1)}$ such that $|X| + |X'| = N$, the states of the form 
\begin{align}
    \left(\prod_{x \in X} \tilde{q}_x^\dag\right) \left(\prod_{x' \in X'} c_{x'}^\dag\right) \ket{\mathrm{vac}}, \label{eq:many-body basis state}
\end{align}
are linearly independent. This can be seen as follows.
Let us assume that 
\begin{align}
    \sum_{X \subset \Lambda^{(1)}} \sum_{X' \subset \Lambda^{(1)}} f(X, X') \left(\prod_{x \in X} \tilde{q}_x^\dag\right) \left(\prod_{x' \in X'} c_{x'}^\dag\right) \ket{\mathrm{vac}} = 0, \label{eq:linear independence of many-body state}
\end{align}
where $f(X,X')$ is a coefficient. 
Since $q_{x, y} = 0$ when $x, y \in \Lambda^{(1)}$, we see that $\{q_x, c_{x'}^\dag\} = 0$ for $x, x' \in \Lambda^{(1)}$.
Keeping this fact, the anticommutation relations \eqref{eq:anticommutation relation 2}, and \eqref{eq:anticommutation relation of q} in mind, we see that operating with
\begin{align}
    \left(\prod_{x \in X} q_x\right) \left(\prod_{x' \in X'} c_{x'}\right)
\end{align}
on Eq.~\eqref{eq:linear independence of many-body state} gives $f(X, X') = 0$.
Thus the states of the form \eqref{eq:many-body basis state} are linearly independent.
The number of states of the form~\eqref{eq:many-body basis state} is counted as follows.
Given that $|X| + |X'| = N$, when $N \leq |\Lambda^{(1)}|$, the number of states is given by
\begin{align}
    \sum_{m=0}^{N} \binom{|\Lambda^{(1)}|}{m} \binom{|\Lambda^{(1)}|}{N-m} = \binom{|\Lambda|}{N}.
\end{align}
When $N > |\Lambda^{(1)}|$, the number of states is also
\begin{align}
    \sum_{m=0}^{|\Lambda|-N} \binom{|\Lambda^{(1)}|}{N - |\Lambda^{(1)}| + m} \binom{|\Lambda^{(1)}|}{|\Lambda^{(1)}|-m} = \binom{|\Lambda|}{N}.
\end{align}
Since these numbers coincide with the dimension of the $N$-particle Hilbert space, we conclude that the states of the form Eq.~\eqref{eq:many-body basis state} span this Hilbert space. 
This immediately implies that  $\ket{\Psi_{\mathrm{GS}}}$, a zero-energy ground state of $H_{\rm par}$, can be expressed as
\begin{align}
    &\ket{\Psi_{\mathrm{GS}}} \nonumber \\
    & = \sum_{X \subset \Lambda^{(1)}} \sum_{X' \subset \Lambda^{(1)}} f_1(X, X') \left(\prod_{x \in X} \tilde{q}_x^\dag\right) \left(\prod_{x' \in X'} c_{x'}^\dag\right) \ket{\mathrm{vac}}, \label{eq:basis1 expression}
\end{align}
where $X$ and $X'$ are subsets of $\Lambda^{(1)}$ such that $|X| + |X'| = N$, and $f_1(X,X')$ is a certain coefficient. 
Similarly,  for subsets $Y, Y' \subset \Lambda^{(2)}$, the states of the form
\begin{align}
    \left(\prod_{y \in Y} c_y^\dag\right) \left(\prod_{y' \in Y'} \tilde{q}_{y'}^\dag\right) \ket{\mathrm{vac}} \label{eq:many-body basis state2}
\end{align}
are linearly independent, and the number of states of the form~\eqref{eq:many-body basis state2} is $\binom{|\Lambda|}{N}$.
Therefore, $\ket{\Psi_{\mathrm{GS}}}$ can also be written as
\begin{align}
    & \ket{\Psi_{\mathrm{GS}}} \nonumber \\ 
    & = \sum_{Y \subset \Lambda^{(2)}} \sum_{Y' \subset \Lambda^{(2)}} f_2(Y, Y') \left(\prod_{y \in Y} c_y^\dag\right) \left(\prod_{y' \in Y'} \tilde{q}_{y'}^\dag\right) \ket{\mathrm{vac}}, \label{eq:basis2 expression}
\end{align}
where $Y$ and $Y'$ are subsets of $\Lambda^{(2)}$ such that $|Y| + |Y'| = N$.
The coefficients $f_1(X, X')$ and $f_2(Y, Y')$ are related as follows:
\begin{align}
    f_1(X, X') 
    = \sum_{\substack{Y \subset \Lambda^{(2)} \\
    |Y| = |X|}}
    \sum_{\substack{Y' \subset \Lambda^{(2)} \\
    |Y'| = |X'|}} \det \mathsf{Q}_{X, Y} \det \mathsf{Q}^{-1}_{X', Y'} f_2(Y, Y'), \label{eq:relation between f1 and f2}
\end{align}
where the matrix $\mathsf{Q}_{X, Y}$ and $\mathsf{Q}_{X, Y}^{-1}$ are $|X|$ by $|Y|$ matrices whose elements are given as $\left(\mathsf{Q}_{X, Y}\right)_{i, j} = \left(\mathsf{Q}\right)_{x_i, y_j}$ and $\left(\mathsf{Q}^{-1}_{X, Y}\right)_{i, j} = \left(\mathsf{Q}^{-1}\right)_{x_i, y_j}$, respectively. Here $x_i$ ($y_j$) denotes the $i$th ($j$th) element of the subset $X$ ($Y$). 
\par
Since $H_{\mathrm{par}}$ is positive-semidefinite, the ground state $\ket{\Psi_{\mathrm{GS}}}$ satisfies $H_{\mathrm{par}}\ket{\Psi_{\mathrm{GS}}} = 0$, which leads to
\begin{align}
    c_z^\dag q_z \ket{\Psi_{\mathrm{GS}}} = 0 \ \ \text{for all} \ z \in \Lambda. \label{eq:cdagq=0}
\end{align}
First, we consider the case of $z \in \Lambda^{(1)}$.
Using the anticommutation relations $\{q_z, c_x^\dag\} = 0$ for all $x, z \in \Lambda^{(1)}$,
we find
\begin{align}
    &\sum_{X, X' \subset \Lambda^{(1)}} f_1(X, X')\, \chi[z \in X]\, \chi[z \notin X'] \nonumber \\ 
    & \times (-1)^{|X|} \mathrm{sgn}(X; z)\, \mathrm{sgn}(X' \cup \{z\}; z)\,   \nonumber \\
    & \times \left( \prod_{x \in X \backslash \{z\}} \tilde{q}_x^\dag\right) 
    \left( \prod_{x' \in X' \cup \{z\}} c_{x'}^\dag\right) \ket{\mathrm{vac}} = 0,
\end{align}
where $\chi[\dots]$ takes the value $1$ if the statement in the brackets is true and $0$ otherwise. 
The function $\mathrm{sgn}(X; z)$, which is a sign factor arising from exchanges of fermion operators, takes $\pm 1$ depending on the position of $z$ in $X$.
Because the states $\left( \prod_{x \in X \backslash \{z\}} \tilde{q}_x^\dag\right) \left( \prod_{x' \in X' \cup \{z\}} c_{x'}^\dag\right) \ket{\mathrm{vac}}$ are linearly independent for all $X$, $X'$, and $z \in \Lambda^{(1)}$, we have 
\begin{align}
    f_1(X, X')\,\chi[z \in X]\, \chi[z \notin X'] = 0,
\end{align}
for all $z \in \Lambda^{(1)}$ and $X, X' \subset \Lambda^{(1)}$ such that $|X| + |X'| = N$, and this implies that 
\begin{align}
    f_1(X, X') = 0 \ \ \text{if} \ \ X \not\subset X'. \label{eq:f1=0}
\end{align}
For $z \in \Lambda^{(2)}$, we repeat the same discussion using the expression~\eqref{eq:basis2 expression}, and then, for subsets $Y, Y' \subset \Lambda^{(2)}$ such that $|Y| + |Y'| = N$, we see that 
\begin{align}
    f_2(Y, Y') = 0 \ \ \text{if} \ \ Y' \not\subset Y. \label{eq:f2=0}
\end{align}
It is clear from Eq.~\eqref{eq:relation between f1 and f2} that $f_1(X, X')$ with $X \subsetneq X'$ can be represented as a sum of $f_2(Y, Y')$ with $|Y| < |Y'|$.
This implies that $Y' \subset Y$ cannot be satisfied, and hence $f_2(Y, Y') = 0$ from Eq.~\eqref{eq:f2=0}.
Therefore, $f_1(X, X') = 0$ when $X \subsetneq X'$. In other words, $f_1(X, X')$ can be nonzero only when $X = X'$.
Therefore, the state $\ket{\Psi_{\mathrm{GS}}}$ can be expressed in the form
\begin{align}
    \ket{\Psi_{\mathrm{GS}}} = \sum_{X \subset \Lambda^{(1)}} f(X) \left(\prod_{x \in X} \tilde{q}_x^\dag\right) 
    \left(\prod_{x' \in X} c_{x'}^\dag\right) \ket{\mathrm{vac}}, \label{eq:sum of only X}
\end{align}
where the sum is over all $X$ such that $2|X| = N$, and we have denoted $f_1(X, X)$ simply as $f(X)$.
We also find that the number of particles in the zero-energy ground states must be an even number.
Hereafter, we assume that $N$ is of the form $N = |\Lambda| - 2k$, where $k = 0, 1,  \dots, |\Lambda|/2$.
\par
In the following, we show that the coefficients satisfy $f(X) = f(X')$; that is, $f(X)$ is independent of $X$ when the number of elements is fixed, which means that the ground state is unique in the $(|\Lambda| - 2k)$-particle Hilbert space.
With Eq.~\eqref{eq:sum of only X}, we examine Eq.~\eqref{eq:cdagq=0} with $z$ replaced by $y \in \Lambda^{(2)}$, and then Eq.~\eqref{eq:cdagq=0} is rewritten as
\begin{align}
    &\sum_{z, z' \in \Lambda^{(1)}} \sum_{X \subset \Lambda^{(1)}} q_{z,y} q_{y, z'} f(X)(-1)^{|X|} \nonumber \\
    & \times \chi[z' \in X]\, \chi[z \not\in X]\, \mathrm{sgn}(X; z')\, \mathrm{sgn}(X \cup \{z\}; z) \nonumber \\ 
    &\times \left(\prod_{x \in X \cup \{z\}} \tilde{q}_x^\dag\right) \left(\prod_{x' \in X \backslash \{z'\}} c_{x'}^\dag\right) \ket{\mathrm{vac}} = 0.
    \label{cdagq=0 for y}
\end{align}
Due to the factor of $\chi[z' \in X]\,\chi[z \not\in X]$, the range of the sum for $X$ is restricted to sets satisfying $z' \in X$ and $z \not\in X$ for a fixed pair $(z, z')$, and such a set $X$ can be written as $X = D \cup \{z'\}$ using a set $D \subset \Lambda^{(1)}$ that does not contain either $z$ or $z'$.
Therefore, we can rewrite Eq.~\eqref{cdagq=0 for y} as
\begin{align}
    & \sum_{z < z'} \sum_{D \subset \Lambda^{(1)}}
    (-1)^{|D|} \mathrm{sgn}(D \cup \{z\}; z)\, \mathrm{sgn}(D \cup \{z', z\}; z') \nonumber \\
    & \times \left[ q_{z, y} q_{y, z'} f(D \cup \{z'\}) - q_{z', y} q_{y, z} f(D \cup \{z\})\right] \nonumber \\
    & \times \left(\prod_{x \in D \cup \{z, z'\}} \tilde{q}_x^\dag\right) 
    \left(\prod_{x' \in D} c_{x'}^\dag\right)
    \ket{\mathrm{vac}} = 0,
\end{align}
which reduces to
\begin{align}
    q_{z, y} q_{y, z'} \left(f(D \cup \{z'\}) - f(D \cup \{z\})\right) = 0,
\end{align}
where $(z, z')$ is an arbitrary pair of different sites in $\Lambda^{(1)}$, and $D$ is a subset of $\Lambda^{(1)}$ such that $z, z' \not \in D$ and $|D| = |\Lambda|/2 - k - 1$.
If $q_{z, y} q_{y, z'} \neq 0$, i.e., $z$ and $z'$ are connected via non-vanishing matrix elements of $\mathsf{Q}$, then $f(D \cup \{z\}) = f(D \cup \{z'\})$.
Since we have assumed that $\mathsf{Q}$ is connected, for an arbitrary $D \subset \Lambda^{(1)}$ such that $|D| = |\Lambda|/2 - k - 1$, we find
\begin{align}
    f(D \cup \{z\}) = f(D \cup \{z'\})
\end{align}
for all $z, z' \in \Lambda^{(1)}$ that are not contained in $D$.
Therefore, letting $X$ be a subset and $X_{z\rightarrow z'}$ be a subset obtained by removing $z$ and adding $z'$ to $X$, we see that
\begin{align}
    f(X) = f(X_{z \rightarrow z'}).
\end{align}
For arbitrary subsets $X$ and $X'$ such that $|X| = |X'|$, we can find pairs $\{z_l, z_l'\}_{l=1}^{n}$ such that $X^{1} = X_{z_1 \rightarrow z_1'}$, $X^2 = X^1_{z_2 \rightarrow z_2'}$, \dots, $X' = X^{n-1}_{z_n \rightarrow z_n'}$.
Thus we obtain $f(X) = f(X')$, which is the desired result.
\par
Finally, we see that the state with $f(X)$ independent of $X$ is indeed the QMBS in Eq.~\eqref{eq:qmbs}.
Now, the ground state is written as 
\begin{align}
    \ket{\Psi_{\mathrm{GS}}} 
    = \sum_{\substack{
    X \subset \Lambda^{(1)} \\
    |X| = |\Lambda|/2 - k
    }} \left(\prod_{x \in X}\tilde{q}_x^\dag c_x^\dag\right)
    \ket{\mathrm{vac}}. \label{eq:ground state qdag cdag}
\end{align}
Because the state $\ket{\mathrm{vac}}$ is proportional to $\left(\prod_{x \in \Lambda^{(1)}}  c_x q_x\right) \ket{\overline{\mathrm{vac}}}$, the ground state Eq.~\eqref{eq:ground state qdag cdag} can be rewritten as
\begin{align}
    \ket{\Psi_{\mathrm{GS}}} 
    \propto \sum_{\substack{
    X \subset \Lambda^{(1)}\\
    |X| = |\Lambda|/2 - k
    }} \left( \prod_{x \in X} \tilde{q}_x^\dag c_x^\dag \right)
    \left(\prod_{x' \in \Lambda^{(1)}}  c_{x'} q_{x'}\right) \ket{\overline{\mathrm{vac}}}.
\end{align}
Using the relations~\eqref{eq:anticommutation relation of q}, we obtain 
\begin{align}
    \ket{\Psi_{\mathrm{GS}}} \propto  \left(\sum_{x, y \in \Lambda} q_{x, y} c_x c_y\right)^k \ket{\overline{\mathrm{vac}}},
\end{align}
which is the QMBS of Eq.~\eqref{eq:qmbs}.
Thus, the unique ground states of the parent Hamiltonian $H_{\rm par}$ are the QMBS.
\bibliographystyle{apsrev4-2}
\bibliography{reference}

\begin{thebibliography}{76}%
\makeatletter
\providecommand \@ifxundefined [1]{%
 \@ifx{#1\undefined}
}%
\providecommand \@ifnum [1]{%
 \ifnum #1\expandafter \@firstoftwo
 \else \expandafter \@secondoftwo
 \fi
}%
\providecommand \@ifx [1]{%
 \ifx #1\expandafter \@firstoftwo
 \else \expandafter \@secondoftwo
 \fi
}%
\providecommand \natexlab [1]{#1}%
\providecommand \enquote  [1]{``#1''}%
\providecommand \bibnamefont  [1]{#1}%
\providecommand \bibfnamefont [1]{#1}%
\providecommand \citenamefont [1]{#1}%
\providecommand \href@noop [0]{\@secondoftwo}%
\providecommand \href [0]{\begingroup \@sanitize@url \@href}%
\providecommand \@href[1]{\@@startlink{#1}\@@href}%
\providecommand \@@href[1]{\endgroup#1\@@endlink}%
\providecommand \@sanitize@url [0]{\catcode `\\12\catcode `\$12\catcode
  `\&12\catcode `\#12\catcode `\^12\catcode `\_12\catcode `\%12\relax}%
\providecommand \@@startlink[1]{}%
\providecommand \@@endlink[0]{}%
\providecommand \url  [0]{\begingroup\@sanitize@url \@url }%
\providecommand \@url [1]{\endgroup\@href {#1}{\urlprefix }}%
\providecommand \urlprefix  [0]{URL }%
\providecommand \Eprint [0]{\href }%
\providecommand \doibase [0]{https://doi.org/}%
\providecommand \selectlanguage [0]{\@gobble}%
\providecommand \bibinfo  [0]{\@secondoftwo}%
\providecommand \bibfield  [0]{\@secondoftwo}%
\providecommand \translation [1]{[#1]}%
\providecommand \BibitemOpen [0]{}%
\providecommand \bibitemStop [0]{}%
\providecommand \bibitemNoStop [0]{.\EOS\space}%
\providecommand \EOS [0]{\spacefactor3000\relax}%
\providecommand \BibitemShut  [1]{\csname bibitem#1\endcsname}%
\let\auto@bib@innerbib\@empty
\bibitem [{\citenamefont {Polkovnikov}\ \emph {et~al.}(2011)\citenamefont
  {Polkovnikov}, \citenamefont {Sengupta}, \citenamefont {Silva},\ and\
  \citenamefont {Vengalattore}}]{polkovnikov2011colloquium}%
  \BibitemOpen
  \bibfield  {author} {\bibinfo {author} {\bibfnamefont {A.}~\bibnamefont
  {Polkovnikov}}, \bibinfo {author} {\bibfnamefont {K.}~\bibnamefont
  {Sengupta}}, \bibinfo {author} {\bibfnamefont {A.}~\bibnamefont {Silva}},\
  and\ \bibinfo {author} {\bibfnamefont {M.}~\bibnamefont {Vengalattore}},\
  }\href {https://doi.org/10.1103/RevModPhys.83.863} {\bibfield  {journal}
  {\bibinfo  {journal} {Rev. Mod. Phys.}\ }\textbf {\bibinfo {volume} {83}},\
  \bibinfo {pages} {863} (\bibinfo {year} {2011})}\BibitemShut {NoStop}%
\bibitem [{\citenamefont {Nandkishore}\ and\ \citenamefont
  {Huse}(2015)}]{nandkishore2015many}%
  \BibitemOpen
  \bibfield  {author} {\bibinfo {author} {\bibfnamefont {R.}~\bibnamefont
  {Nandkishore}}\ and\ \bibinfo {author} {\bibfnamefont {D.~A.}\ \bibnamefont
  {Huse}},\ }\href {https://doi.org/10.1146/annurev-conmatphys-031214-014726}
  {\bibfield  {journal} {\bibinfo  {journal} {Annu. Rev. Condens. Matter
  Phys.}\ }\textbf {\bibinfo {volume} {6}},\ \bibinfo {pages} {15} (\bibinfo
  {year} {2015})}\BibitemShut {NoStop}%
\bibitem [{\citenamefont {Deutsch}(1991)}]{deutsch1991quantum}%
  \BibitemOpen
  \bibfield  {author} {\bibinfo {author} {\bibfnamefont {J.~M.}\ \bibnamefont
  {Deutsch}},\ }\href {https://doi.org/10.1103/PhysRevA.43.2046} {\bibfield
  {journal} {\bibinfo  {journal} {Phys. Rev. A}\ }\textbf {\bibinfo {volume}
  {43}},\ \bibinfo {pages} {2046} (\bibinfo {year} {1991})}\BibitemShut
  {NoStop}%
\bibitem [{\citenamefont {Srednicki}(1994)}]{srednicki1994chaos}%
  \BibitemOpen
  \bibfield  {author} {\bibinfo {author} {\bibfnamefont {M.}~\bibnamefont
  {Srednicki}},\ }\href {https://doi.org/10.1103/PhysRevE.50.888} {\bibfield
  {journal} {\bibinfo  {journal} {Phys. Rev. E}\ }\textbf {\bibinfo {volume}
  {50}},\ \bibinfo {pages} {888} (\bibinfo {year} {1994})}\BibitemShut
  {NoStop}%
\bibitem [{\citenamefont {Rigol}\ \emph {et~al.}(2008)\citenamefont {Rigol},
  \citenamefont {Dunjko},\ and\ \citenamefont
  {Olshanii}}]{rigol2008thermalization}%
  \BibitemOpen
  \bibfield  {author} {\bibinfo {author} {\bibfnamefont {M.}~\bibnamefont
  {Rigol}}, \bibinfo {author} {\bibfnamefont {V.}~\bibnamefont {Dunjko}},\ and\
  \bibinfo {author} {\bibfnamefont {M.}~\bibnamefont {Olshanii}},\ }\href
  {https://doi.org/10.1038/nature06838} {\bibfield  {journal} {\bibinfo
  {journal} {Nature}\ }\textbf {\bibinfo {volume} {452}},\ \bibinfo {pages}
  {854} (\bibinfo {year} {2008})}\BibitemShut {NoStop}%
\bibitem [{\citenamefont {Kim}\ \emph {et~al.}(2014)\citenamefont {Kim},
  \citenamefont {Ikeda},\ and\ \citenamefont {Huse}}]{kim2014testing}%
  \BibitemOpen
  \bibfield  {author} {\bibinfo {author} {\bibfnamefont {H.}~\bibnamefont
  {Kim}}, \bibinfo {author} {\bibfnamefont {T.~N.}\ \bibnamefont {Ikeda}},\
  and\ \bibinfo {author} {\bibfnamefont {D.~A.}\ \bibnamefont {Huse}},\ }\href
  {https://doi.org/10.1103/PhysRevE.90.052105} {\bibfield  {journal} {\bibinfo
  {journal} {Phys. Rev. E}\ }\textbf {\bibinfo {volume} {90}},\ \bibinfo
  {pages} {052105} (\bibinfo {year} {2014})}\BibitemShut {NoStop}%
\bibitem [{\citenamefont {D'Alessio}\ \emph {et~al.}(2016)\citenamefont
  {D'Alessio}, \citenamefont {Kafri}, \citenamefont {Polkovnikov},\ and\
  \citenamefont {Rigol}}]{d2016quantum}%
  \BibitemOpen
  \bibfield  {author} {\bibinfo {author} {\bibfnamefont {L.}~\bibnamefont
  {D'Alessio}}, \bibinfo {author} {\bibfnamefont {Y.}~\bibnamefont {Kafri}},
  \bibinfo {author} {\bibfnamefont {A.}~\bibnamefont {Polkovnikov}},\ and\
  \bibinfo {author} {\bibfnamefont {M.}~\bibnamefont {Rigol}},\ }\href
  {https://doi.org/10.1080/00018732.2016.1198134} {\bibfield  {journal}
  {\bibinfo  {journal} {Adv. Phys.}\ }\textbf {\bibinfo {volume} {65}},\
  \bibinfo {pages} {239} (\bibinfo {year} {2016})}\BibitemShut {NoStop}%
\bibitem [{\citenamefont {Bernien}\ \emph {et~al.}(2017)\citenamefont
  {Bernien}, \citenamefont {Schwartz}, \citenamefont {Keesling}, \citenamefont
  {Levine}, \citenamefont {Omran}, \citenamefont {Pichler}, \citenamefont
  {Choi}, \citenamefont {Zibrov}, \citenamefont {Endres}, \citenamefont
  {Greiner} \emph {et~al.}}]{bernien2017probing}%
  \BibitemOpen
  \bibfield  {author} {\bibinfo {author} {\bibfnamefont {H.}~\bibnamefont
  {Bernien}}, \bibinfo {author} {\bibfnamefont {S.}~\bibnamefont {Schwartz}},
  \bibinfo {author} {\bibfnamefont {A.}~\bibnamefont {Keesling}}, \bibinfo
  {author} {\bibfnamefont {H.}~\bibnamefont {Levine}}, \bibinfo {author}
  {\bibfnamefont {A.}~\bibnamefont {Omran}}, \bibinfo {author} {\bibfnamefont
  {H.}~\bibnamefont {Pichler}}, \bibinfo {author} {\bibfnamefont
  {S.}~\bibnamefont {Choi}}, \bibinfo {author} {\bibfnamefont {A.~S.}\
  \bibnamefont {Zibrov}}, \bibinfo {author} {\bibfnamefont {M.}~\bibnamefont
  {Endres}}, \bibinfo {author} {\bibfnamefont {M.}~\bibnamefont {Greiner}},
  \emph {et~al.},\ }\href {https://doi.org/10.1038/nature24622} {\bibfield
  {journal} {\bibinfo  {journal} {Nature}\ }\textbf {\bibinfo {volume} {551}},\
  \bibinfo {pages} {579} (\bibinfo {year} {2017})}\BibitemShut {NoStop}%
\bibitem [{\citenamefont {Su}\ \emph {et~al.}(2022)\citenamefont {Su},
  \citenamefont {Sun}, \citenamefont {Hudomal}, \citenamefont {Desaules},
  \citenamefont {Zhou}, \citenamefont {Yang}, \citenamefont {Halimeh},
  \citenamefont {Yuan}, \citenamefont {Papi{\'c}},\ and\ \citenamefont
  {Pan}}]{su2022observation}%
  \BibitemOpen
  \bibfield  {author} {\bibinfo {author} {\bibfnamefont {G.-X.}\ \bibnamefont
  {Su}}, \bibinfo {author} {\bibfnamefont {H.}~\bibnamefont {Sun}}, \bibinfo
  {author} {\bibfnamefont {A.}~\bibnamefont {Hudomal}}, \bibinfo {author}
  {\bibfnamefont {J.-Y.}\ \bibnamefont {Desaules}}, \bibinfo {author}
  {\bibfnamefont {Z.-Y.}\ \bibnamefont {Zhou}}, \bibinfo {author}
  {\bibfnamefont {B.}~\bibnamefont {Yang}}, \bibinfo {author} {\bibfnamefont
  {J.~C.}\ \bibnamefont {Halimeh}}, \bibinfo {author} {\bibfnamefont {Z.-S.}\
  \bibnamefont {Yuan}}, \bibinfo {author} {\bibfnamefont {Z.}~\bibnamefont
  {Papi{\'c}}},\ and\ \bibinfo {author} {\bibfnamefont {J.-W.}\ \bibnamefont
  {Pan}},\ }\href {https://arxiv.org/abs/2201.00821} {\bibfield  {journal}
  {\bibinfo  {journal} {arXiv:2201.00821}\ } (\bibinfo {year}
  {2022})}\BibitemShut {NoStop}%
\bibitem [{\citenamefont {Serbyn}\ \emph {et~al.}(2021)\citenamefont {Serbyn},
  \citenamefont {Abanin},\ and\ \citenamefont {Papi{\'c}}}]{serbyn2021quantum}%
  \BibitemOpen
  \bibfield  {author} {\bibinfo {author} {\bibfnamefont {M.}~\bibnamefont
  {Serbyn}}, \bibinfo {author} {\bibfnamefont {D.~A.}\ \bibnamefont {Abanin}},\
  and\ \bibinfo {author} {\bibfnamefont {Z.}~\bibnamefont {Papi{\'c}}},\ }\href
  {https://doi.org/10.1038/s41567-021-01230-2} {\bibfield  {journal} {\bibinfo
  {journal} {Nat. Phys.}\ }\textbf {\bibinfo {volume} {17}},\ \bibinfo {pages}
  {675} (\bibinfo {year} {2021})}\BibitemShut {NoStop}%
\bibitem [{\citenamefont {Regnault}\ \emph {et~al.}(2022)\citenamefont
  {Regnault}, \citenamefont {Moudgalya},\ and\ \citenamefont
  {Bernevig}}]{regnault2022quantum}%
  \BibitemOpen
  \bibfield  {author} {\bibinfo {author} {\bibfnamefont {N.}~\bibnamefont
  {Regnault}}, \bibinfo {author} {\bibfnamefont {S.}~\bibnamefont
  {Moudgalya}},\ and\ \bibinfo {author} {\bibfnamefont {B.~A.}\ \bibnamefont
  {Bernevig}},\ }\href {https://doi.org/10.1088/1361-6633/ac73a0} {\bibfield
  {journal} {\bibinfo  {journal} {Rep. Prog. Phys.}\ }\textbf {\bibinfo
  {volume} {85}},\ \bibinfo {pages} {086501} (\bibinfo {year}
  {2022})}\BibitemShut {NoStop}%
\bibitem [{\citenamefont {Chandran}\ \emph {et~al.}(2022)\citenamefont
  {Chandran}, \citenamefont {Iadecola}, \citenamefont {Khemani},\ and\
  \citenamefont {Moessner}}]{chandran2022quantum}%
  \BibitemOpen
  \bibfield  {author} {\bibinfo {author} {\bibfnamefont {A.}~\bibnamefont
  {Chandran}}, \bibinfo {author} {\bibfnamefont {T.}~\bibnamefont {Iadecola}},
  \bibinfo {author} {\bibfnamefont {V.}~\bibnamefont {Khemani}},\ and\ \bibinfo
  {author} {\bibfnamefont {R.}~\bibnamefont {Moessner}},\ }\href
  {https://doi.org/10.48550/arXiv.2206.11528} {\bibfield  {journal} {\bibinfo
  {journal} {arXiv:2206.11528}\ } (\bibinfo {year} {2022})}\BibitemShut
  {NoStop}%
\bibitem [{\citenamefont {Papi{\'c}}(2021)}]{papic2021weak}%
  \BibitemOpen
  \bibfield  {author} {\bibinfo {author} {\bibfnamefont {Z.}~\bibnamefont
  {Papi{\'c}}},\ }\href {https://arxiv.org/abs/2108.03460} {\bibfield
  {journal} {\bibinfo  {journal} {arXiv:2108.03460}\ } (\bibinfo {year}
  {2021})}\BibitemShut {NoStop}%
\bibitem [{\citenamefont {Choi}\ \emph {et~al.}(2019)\citenamefont {Choi},
  \citenamefont {Turner}, \citenamefont {Pichler}, \citenamefont {Ho},
  \citenamefont {Michailidis}, \citenamefont {Papi{\'c}}, \citenamefont
  {Serbyn}, \citenamefont {Lukin},\ and\ \citenamefont
  {Abanin}}]{choi2019emergent}%
  \BibitemOpen
  \bibfield  {author} {\bibinfo {author} {\bibfnamefont {S.}~\bibnamefont
  {Choi}}, \bibinfo {author} {\bibfnamefont {C.~J.}\ \bibnamefont {Turner}},
  \bibinfo {author} {\bibfnamefont {H.}~\bibnamefont {Pichler}}, \bibinfo
  {author} {\bibfnamefont {W.~W.}\ \bibnamefont {Ho}}, \bibinfo {author}
  {\bibfnamefont {A.~A.}\ \bibnamefont {Michailidis}}, \bibinfo {author}
  {\bibfnamefont {Z.}~\bibnamefont {Papi{\'c}}}, \bibinfo {author}
  {\bibfnamefont {M.}~\bibnamefont {Serbyn}}, \bibinfo {author} {\bibfnamefont
  {M.~D.}\ \bibnamefont {Lukin}},\ and\ \bibinfo {author} {\bibfnamefont
  {D.~A.}\ \bibnamefont {Abanin}},\ }\href
  {https://doi.org/10.1103/PhysRevLett.122.220603} {\bibfield  {journal}
  {\bibinfo  {journal} {Phys. Rev. Lett.}\ }\textbf {\bibinfo {volume} {122}},\
  \bibinfo {pages} {220603} (\bibinfo {year} {2019})}\BibitemShut {NoStop}%
\bibitem [{\citenamefont {Ho}\ \emph {et~al.}(2019)\citenamefont {Ho},
  \citenamefont {Choi}, \citenamefont {Pichler},\ and\ \citenamefont
  {Lukin}}]{ho2019periodic}%
  \BibitemOpen
  \bibfield  {author} {\bibinfo {author} {\bibfnamefont {W.~W.}\ \bibnamefont
  {Ho}}, \bibinfo {author} {\bibfnamefont {S.}~\bibnamefont {Choi}}, \bibinfo
  {author} {\bibfnamefont {H.}~\bibnamefont {Pichler}},\ and\ \bibinfo {author}
  {\bibfnamefont {M.~D.}\ \bibnamefont {Lukin}},\ }\href
  {https://doi.org/10.1103/PhysRevLett.122.040603} {\bibfield  {journal}
  {\bibinfo  {journal} {Phys. Rev. Lett.}\ }\textbf {\bibinfo {volume} {122}},\
  \bibinfo {pages} {040603} (\bibinfo {year} {2019})}\BibitemShut {NoStop}%
\bibitem [{\citenamefont {Lin}\ and\ \citenamefont
  {Motrunich}(2019)}]{lin2019exact}%
  \BibitemOpen
  \bibfield  {author} {\bibinfo {author} {\bibfnamefont {C.-J.}\ \bibnamefont
  {Lin}}\ and\ \bibinfo {author} {\bibfnamefont {O.~I.}\ \bibnamefont
  {Motrunich}},\ }\href {https://doi.org/10.1103/PhysRevLett.122.173401}
  {\bibfield  {journal} {\bibinfo  {journal} {Phys. Rev. Lett.}\ }\textbf
  {\bibinfo {volume} {122}},\ \bibinfo {pages} {173401} (\bibinfo {year}
  {2019})}\BibitemShut {NoStop}%
\bibitem [{\citenamefont {Khemani}\ \emph {et~al.}(2019)\citenamefont
  {Khemani}, \citenamefont {Laumann},\ and\ \citenamefont
  {Chandran}}]{khemani2019signatures}%
  \BibitemOpen
  \bibfield  {author} {\bibinfo {author} {\bibfnamefont {V.}~\bibnamefont
  {Khemani}}, \bibinfo {author} {\bibfnamefont {C.~R.}\ \bibnamefont
  {Laumann}},\ and\ \bibinfo {author} {\bibfnamefont {A.}~\bibnamefont
  {Chandran}},\ }\href {https://doi.org/10.1103/PhysRevB.99.161101} {\bibfield
  {journal} {\bibinfo  {journal} {Phys. Rev. B}\ }\textbf {\bibinfo {volume}
  {99}},\ \bibinfo {pages} {161101(R)} (\bibinfo {year} {2019})}\BibitemShut
  {NoStop}%
\bibitem [{\citenamefont {Iadecola}\ \emph {et~al.}(2019)\citenamefont
  {Iadecola}, \citenamefont {Schecter},\ and\ \citenamefont
  {Xu}}]{iadecola2019quantum}%
  \BibitemOpen
  \bibfield  {author} {\bibinfo {author} {\bibfnamefont {T.}~\bibnamefont
  {Iadecola}}, \bibinfo {author} {\bibfnamefont {M.}~\bibnamefont {Schecter}},\
  and\ \bibinfo {author} {\bibfnamefont {S.}~\bibnamefont {Xu}},\ }\href
  {https://doi.org/10.1103/PhysRevB.100.184312} {\bibfield  {journal} {\bibinfo
   {journal} {Phys. Rev. B}\ }\textbf {\bibinfo {volume} {100}},\ \bibinfo
  {pages} {184312} (\bibinfo {year} {2019})}\BibitemShut {NoStop}%
\bibitem [{\citenamefont {Desaules}\ \emph {et~al.}(2022)\citenamefont
  {Desaules}, \citenamefont {Banerjee}, \citenamefont {Hudomal}, \citenamefont
  {Papi{\'c}}, \citenamefont {Sen},\ and\ \citenamefont
  {Halimeh}}]{desaules2022weak}%
  \BibitemOpen
  \bibfield  {author} {\bibinfo {author} {\bibfnamefont {J.-Y.}\ \bibnamefont
  {Desaules}}, \bibinfo {author} {\bibfnamefont {D.}~\bibnamefont {Banerjee}},
  \bibinfo {author} {\bibfnamefont {A.}~\bibnamefont {Hudomal}}, \bibinfo
  {author} {\bibfnamefont {Z.}~\bibnamefont {Papi{\'c}}}, \bibinfo {author}
  {\bibfnamefont {A.}~\bibnamefont {Sen}},\ and\ \bibinfo {author}
  {\bibfnamefont {J.~C.}\ \bibnamefont {Halimeh}},\ }\href
  {https://arxiv.org/abs/2203.08830} {\bibfield  {journal} {\bibinfo  {journal}
  {arXiv:2203.08830}\ } (\bibinfo {year} {2022})}\BibitemShut {NoStop}%
\bibitem [{\citenamefont {Shiraishi}\ and\ \citenamefont
  {Mori}(2017)}]{shiraishi2017systematic}%
  \BibitemOpen
  \bibfield  {author} {\bibinfo {author} {\bibfnamefont {N.}~\bibnamefont
  {Shiraishi}}\ and\ \bibinfo {author} {\bibfnamefont {T.}~\bibnamefont
  {Mori}},\ }\href {https://doi.org/10.1103/PhysRevLett.119.030601} {\bibfield
  {journal} {\bibinfo  {journal} {Phys. Rev. Lett.}\ }\textbf {\bibinfo
  {volume} {119}},\ \bibinfo {pages} {030601} (\bibinfo {year}
  {2017})}\BibitemShut {NoStop}%
\bibitem [{\citenamefont {Moudgalya}\ \emph {et~al.}(2018)\citenamefont
  {Moudgalya}, \citenamefont {Regnault},\ and\ \citenamefont
  {Bernevig}}]{moudgalya2018entanglement}%
  \BibitemOpen
  \bibfield  {author} {\bibinfo {author} {\bibfnamefont {S.}~\bibnamefont
  {Moudgalya}}, \bibinfo {author} {\bibfnamefont {N.}~\bibnamefont
  {Regnault}},\ and\ \bibinfo {author} {\bibfnamefont {B.~A.}\ \bibnamefont
  {Bernevig}},\ }\href {https://doi.org/10.1103/PhysRevB.98.235156} {\bibfield
  {journal} {\bibinfo  {journal} {Phys. Rev. B}\ }\textbf {\bibinfo {volume}
  {98}},\ \bibinfo {pages} {235156} (\bibinfo {year} {2018})}\BibitemShut
  {NoStop}%
\bibitem [{\citenamefont {Pai}\ and\ \citenamefont
  {Pretko}(2019)}]{pai2019dynamical}%
  \BibitemOpen
  \bibfield  {author} {\bibinfo {author} {\bibfnamefont {S.}~\bibnamefont
  {Pai}}\ and\ \bibinfo {author} {\bibfnamefont {M.}~\bibnamefont {Pretko}},\
  }\href {https://doi.org/10.1103/PhysRevLett.123.136401} {\bibfield  {journal}
  {\bibinfo  {journal} {Phys. Rev. Lett.}\ }\textbf {\bibinfo {volume} {123}},\
  \bibinfo {pages} {136401} (\bibinfo {year} {2019})}\BibitemShut {NoStop}%
\bibitem [{\citenamefont {Chattopadhyay}\ \emph {et~al.}(2020)\citenamefont
  {Chattopadhyay}, \citenamefont {Pichler}, \citenamefont {Lukin},\ and\
  \citenamefont {Ho}}]{chattopadhyay2020quantum}%
  \BibitemOpen
  \bibfield  {author} {\bibinfo {author} {\bibfnamefont {S.}~\bibnamefont
  {Chattopadhyay}}, \bibinfo {author} {\bibfnamefont {H.}~\bibnamefont
  {Pichler}}, \bibinfo {author} {\bibfnamefont {M.~D.}\ \bibnamefont {Lukin}},\
  and\ \bibinfo {author} {\bibfnamefont {W.~W.}\ \bibnamefont {Ho}},\ }\href
  {https://doi.org/10.1103/PhysRevB.101.174308} {\bibfield  {journal} {\bibinfo
   {journal} {Phys. Rev. B}\ }\textbf {\bibinfo {volume} {101}},\ \bibinfo
  {pages} {174308} (\bibinfo {year} {2020})}\BibitemShut {NoStop}%
\bibitem [{\citenamefont {Moudgalya}\ \emph
  {et~al.}(2020{\natexlab{a}})\citenamefont {Moudgalya}, \citenamefont
  {Bernevig},\ and\ \citenamefont {Regnault}}]{moudgalya2020quantum}%
  \BibitemOpen
  \bibfield  {author} {\bibinfo {author} {\bibfnamefont {S.}~\bibnamefont
  {Moudgalya}}, \bibinfo {author} {\bibfnamefont {B.~A.}\ \bibnamefont
  {Bernevig}},\ and\ \bibinfo {author} {\bibfnamefont {N.}~\bibnamefont
  {Regnault}},\ }\href {https://doi.org/10.1103/PhysRevB.102.195150} {\bibfield
   {journal} {\bibinfo  {journal} {Phys. Rev. B}\ }\textbf {\bibinfo {volume}
  {102}},\ \bibinfo {pages} {195150} (\bibinfo {year}
  {2020}{\natexlab{a}})}\BibitemShut {NoStop}%
\bibitem [{\citenamefont {Pakrouski}\ \emph {et~al.}(2020)\citenamefont
  {Pakrouski}, \citenamefont {Pallegar}, \citenamefont {Popov},\ and\
  \citenamefont {Klebanov}}]{pakrouski2020many}%
  \BibitemOpen
  \bibfield  {author} {\bibinfo {author} {\bibfnamefont {K.}~\bibnamefont
  {Pakrouski}}, \bibinfo {author} {\bibfnamefont {P.~N.}\ \bibnamefont
  {Pallegar}}, \bibinfo {author} {\bibfnamefont {F.~K.}\ \bibnamefont
  {Popov}},\ and\ \bibinfo {author} {\bibfnamefont {I.~R.}\ \bibnamefont
  {Klebanov}},\ }\href {https://doi.org/10.1103/PhysRevLett.125.230602}
  {\bibfield  {journal} {\bibinfo  {journal} {Phys. Rev. Lett.}\ }\textbf
  {\bibinfo {volume} {125}},\ \bibinfo {pages} {230602} (\bibinfo {year}
  {2020})}\BibitemShut {NoStop}%
\bibitem [{\citenamefont {Moudgalya}\ \emph
  {et~al.}(2020{\natexlab{b}})\citenamefont {Moudgalya}, \citenamefont
  {O'Brien}, \citenamefont {Bernevig}, \citenamefont {Fendley},\ and\
  \citenamefont {Regnault}}]{moudgalya2020large}%
  \BibitemOpen
  \bibfield  {author} {\bibinfo {author} {\bibfnamefont {S.}~\bibnamefont
  {Moudgalya}}, \bibinfo {author} {\bibfnamefont {E.}~\bibnamefont {O'Brien}},
  \bibinfo {author} {\bibfnamefont {B.~A.}\ \bibnamefont {Bernevig}}, \bibinfo
  {author} {\bibfnamefont {P.}~\bibnamefont {Fendley}},\ and\ \bibinfo {author}
  {\bibfnamefont {N.}~\bibnamefont {Regnault}},\ }\href
  {https://doi.org/10.1103/PhysRevB.102.085120} {\bibfield  {journal} {\bibinfo
   {journal} {Phys. Rev. B}\ }\textbf {\bibinfo {volume} {102}},\ \bibinfo
  {pages} {085120} (\bibinfo {year} {2020}{\natexlab{b}})}\BibitemShut
  {NoStop}%
\bibitem [{\citenamefont {Iadecola}\ and\ \citenamefont
  {Schecter}(2020)}]{iadecola2020quantum}%
  \BibitemOpen
  \bibfield  {author} {\bibinfo {author} {\bibfnamefont {T.}~\bibnamefont
  {Iadecola}}\ and\ \bibinfo {author} {\bibfnamefont {M.}~\bibnamefont
  {Schecter}},\ }\href {https://doi.org/10.1103/PhysRevB.101.024306} {\bibfield
   {journal} {\bibinfo  {journal} {Phys. Rev. B}\ }\textbf {\bibinfo {volume}
  {101}},\ \bibinfo {pages} {024306} (\bibinfo {year} {2020})}\BibitemShut
  {NoStop}%
\bibitem [{\citenamefont {Kuno}\ \emph {et~al.}(2020)\citenamefont {Kuno},
  \citenamefont {Mizoguchi},\ and\ \citenamefont {Hatsugai}}]{kuno2020flat}%
  \BibitemOpen
  \bibfield  {author} {\bibinfo {author} {\bibfnamefont {Y.}~\bibnamefont
  {Kuno}}, \bibinfo {author} {\bibfnamefont {T.}~\bibnamefont {Mizoguchi}},\
  and\ \bibinfo {author} {\bibfnamefont {Y.}~\bibnamefont {Hatsugai}},\ }\href
  {https://doi.org/10.1103/PhysRevB.102.241115} {\bibfield  {journal} {\bibinfo
   {journal} {Phys. Rev. B}\ }\textbf {\bibinfo {volume} {102}},\ \bibinfo
  {pages} {241115(R)} (\bibinfo {year} {2020})}\BibitemShut {NoStop}%
\bibitem [{\citenamefont {Sugiura}\ \emph {et~al.}(2021)\citenamefont
  {Sugiura}, \citenamefont {Kuwahara},\ and\ \citenamefont
  {Saito}}]{sugiura2021many}%
  \BibitemOpen
  \bibfield  {author} {\bibinfo {author} {\bibfnamefont {S.}~\bibnamefont
  {Sugiura}}, \bibinfo {author} {\bibfnamefont {T.}~\bibnamefont {Kuwahara}},\
  and\ \bibinfo {author} {\bibfnamefont {K.}~\bibnamefont {Saito}},\ }\href
  {https://doi.org/10.1103/PhysRevResearch.3.L012010} {\bibfield  {journal}
  {\bibinfo  {journal} {Phys. Rev. Res.}\ }\textbf {\bibinfo {volume} {3}},\
  \bibinfo {pages} {L012010} (\bibinfo {year} {2021})}\BibitemShut {NoStop}%
\bibitem [{\citenamefont {Langlett}\ \emph {et~al.}(2022)\citenamefont
  {Langlett}, \citenamefont {Yang}, \citenamefont {Wildeboer}, \citenamefont
  {Gorshkov}, \citenamefont {Iadecola},\ and\ \citenamefont
  {Xu}}]{langlett2022rainbow}%
  \BibitemOpen
  \bibfield  {author} {\bibinfo {author} {\bibfnamefont {C.~M.}\ \bibnamefont
  {Langlett}}, \bibinfo {author} {\bibfnamefont {Z.-C.}\ \bibnamefont {Yang}},
  \bibinfo {author} {\bibfnamefont {J.}~\bibnamefont {Wildeboer}}, \bibinfo
  {author} {\bibfnamefont {A.~V.}\ \bibnamefont {Gorshkov}}, \bibinfo {author}
  {\bibfnamefont {T.}~\bibnamefont {Iadecola}},\ and\ \bibinfo {author}
  {\bibfnamefont {S.}~\bibnamefont {Xu}},\ }\href
  {https://doi.org/10.1103/PhysRevB.105.L060301} {\bibfield  {journal}
  {\bibinfo  {journal} {Phys. Rev. B}\ }\textbf {\bibinfo {volume} {105}},\
  \bibinfo {pages} {L060301} (\bibinfo {year} {2022})}\BibitemShut {NoStop}%
\bibitem [{\citenamefont {Schecter}\ and\ \citenamefont
  {Iadecola}(2019)}]{schecter2019weak}%
  \BibitemOpen
  \bibfield  {author} {\bibinfo {author} {\bibfnamefont {M.}~\bibnamefont
  {Schecter}}\ and\ \bibinfo {author} {\bibfnamefont {T.}~\bibnamefont
  {Iadecola}},\ }\href {https://doi.org/10.1103/PhysRevLett.123.147201}
  {\bibfield  {journal} {\bibinfo  {journal} {Phys. Rev. Lett.}\ }\textbf
  {\bibinfo {volume} {123}},\ \bibinfo {pages} {147201} (\bibinfo {year}
  {2019})}\BibitemShut {NoStop}%
\bibitem [{\citenamefont {Ok}\ \emph {et~al.}(2019)\citenamefont {Ok},
  \citenamefont {Choo}, \citenamefont {Mudry}, \citenamefont {Castelnovo},
  \citenamefont {Chamon},\ and\ \citenamefont {Neupert}}]{ok2019topological}%
  \BibitemOpen
  \bibfield  {author} {\bibinfo {author} {\bibfnamefont {S.}~\bibnamefont
  {Ok}}, \bibinfo {author} {\bibfnamefont {K.}~\bibnamefont {Choo}}, \bibinfo
  {author} {\bibfnamefont {C.}~\bibnamefont {Mudry}}, \bibinfo {author}
  {\bibfnamefont {C.}~\bibnamefont {Castelnovo}}, \bibinfo {author}
  {\bibfnamefont {C.}~\bibnamefont {Chamon}},\ and\ \bibinfo {author}
  {\bibfnamefont {T.}~\bibnamefont {Neupert}},\ }\href
  {https://doi.org/10.1103/PhysRevResearch.1.033144} {\bibfield  {journal}
  {\bibinfo  {journal} {Phys. Rev. Res.}\ }\textbf {\bibinfo {volume} {1}},\
  \bibinfo {pages} {033144} (\bibinfo {year} {2019})}\BibitemShut {NoStop}%
\bibitem [{\citenamefont {Lee}\ \emph {et~al.}(2020)\citenamefont {Lee},
  \citenamefont {Melendrez}, \citenamefont {Pal},\ and\ \citenamefont
  {Changlani}}]{lee2020exact}%
  \BibitemOpen
  \bibfield  {author} {\bibinfo {author} {\bibfnamefont {K.}~\bibnamefont
  {Lee}}, \bibinfo {author} {\bibfnamefont {R.}~\bibnamefont {Melendrez}},
  \bibinfo {author} {\bibfnamefont {A.}~\bibnamefont {Pal}},\ and\ \bibinfo
  {author} {\bibfnamefont {H.~J.}\ \bibnamefont {Changlani}},\ }\href
  {https://doi.org/10.1103/PhysRevB.101.241111} {\bibfield  {journal} {\bibinfo
   {journal} {Phys. Rev. B}\ }\textbf {\bibinfo {volume} {101}},\ \bibinfo
  {pages} {241111(R)} (\bibinfo {year} {2020})}\BibitemShut {NoStop}%
\bibitem [{\citenamefont {Surace}\ \emph {et~al.}(2020)\citenamefont {Surace},
  \citenamefont {Giudici},\ and\ \citenamefont {Dalmonte}}]{surace2020weak}%
  \BibitemOpen
  \bibfield  {author} {\bibinfo {author} {\bibfnamefont {F.~M.}\ \bibnamefont
  {Surace}}, \bibinfo {author} {\bibfnamefont {G.}~\bibnamefont {Giudici}},\
  and\ \bibinfo {author} {\bibfnamefont {M.}~\bibnamefont {Dalmonte}},\ }\href
  {https://doi.org/10.22331/q-2020-10-07-339} {\bibfield  {journal} {\bibinfo
  {journal} {Quantum}\ }\textbf {\bibinfo {volume} {4}},\ \bibinfo {pages}
  {339} (\bibinfo {year} {2020})}\BibitemShut {NoStop}%
\bibitem [{\citenamefont {Michailidis}\ \emph {et~al.}(2020)\citenamefont
  {Michailidis}, \citenamefont {Turner}, \citenamefont {Papi{\'c}},
  \citenamefont {Abanin},\ and\ \citenamefont
  {Serbyn}}]{michailidis2020stabilizing}%
  \BibitemOpen
  \bibfield  {author} {\bibinfo {author} {\bibfnamefont {A.~A.}\ \bibnamefont
  {Michailidis}}, \bibinfo {author} {\bibfnamefont {C.~J.}\ \bibnamefont
  {Turner}}, \bibinfo {author} {\bibfnamefont {Z.}~\bibnamefont {Papi{\'c}}},
  \bibinfo {author} {\bibfnamefont {D.~A.}\ \bibnamefont {Abanin}},\ and\
  \bibinfo {author} {\bibfnamefont {M.}~\bibnamefont {Serbyn}},\ }\href
  {https://doi.org/10.1103/PhysRevResearch.2.022065} {\bibfield  {journal}
  {\bibinfo  {journal} {Phys. Rev. Res.}\ }\textbf {\bibinfo {volume} {2}},\
  \bibinfo {pages} {022065(R)} (\bibinfo {year} {2020})}\BibitemShut {NoStop}%
\bibitem [{\citenamefont {Lin}\ \emph {et~al.}(2020)\citenamefont {Lin},
  \citenamefont {Calvera},\ and\ \citenamefont {Hsieh}}]{lin2020quantum}%
  \BibitemOpen
  \bibfield  {author} {\bibinfo {author} {\bibfnamefont {C.-J.}\ \bibnamefont
  {Lin}}, \bibinfo {author} {\bibfnamefont {V.}~\bibnamefont {Calvera}},\ and\
  \bibinfo {author} {\bibfnamefont {T.~H.}\ \bibnamefont {Hsieh}},\ }\href
  {https://doi.org/10.1103/PhysRevB.101.220304} {\bibfield  {journal} {\bibinfo
   {journal} {Phys. Rev. B}\ }\textbf {\bibinfo {volume} {101}},\ \bibinfo
  {pages} {220304(R)} (\bibinfo {year} {2020})}\BibitemShut {NoStop}%
\bibitem [{\citenamefont {Wildeboer}\ \emph {et~al.}(2021)\citenamefont
  {Wildeboer}, \citenamefont {Seidel}, \citenamefont {Srivatsa}, \citenamefont
  {Nielsen},\ and\ \citenamefont {Erten}}]{wildeboer2021topological}%
  \BibitemOpen
  \bibfield  {author} {\bibinfo {author} {\bibfnamefont {J.}~\bibnamefont
  {Wildeboer}}, \bibinfo {author} {\bibfnamefont {A.}~\bibnamefont {Seidel}},
  \bibinfo {author} {\bibfnamefont {N.~S.}\ \bibnamefont {Srivatsa}}, \bibinfo
  {author} {\bibfnamefont {A.~E.~B.}\ \bibnamefont {Nielsen}},\ and\ \bibinfo
  {author} {\bibfnamefont {O.}~\bibnamefont {Erten}},\ }\href
  {https://doi.org/10.1103/PhysRevB.104.L121103} {\bibfield  {journal}
  {\bibinfo  {journal} {Phys. Rev. B}\ }\textbf {\bibinfo {volume} {104}},\
  \bibinfo {pages} {L121103} (\bibinfo {year} {2021})}\BibitemShut {NoStop}%
\bibitem [{\citenamefont {McClarty}\ \emph {et~al.}(2020)\citenamefont
  {McClarty}, \citenamefont {Haque}, \citenamefont {Sen},\ and\ \citenamefont
  {Richter}}]{PhysRevB.102.224303}%
  \BibitemOpen
  \bibfield  {author} {\bibinfo {author} {\bibfnamefont {P.~A.}\ \bibnamefont
  {McClarty}}, \bibinfo {author} {\bibfnamefont {M.}~\bibnamefont {Haque}},
  \bibinfo {author} {\bibfnamefont {A.}~\bibnamefont {Sen}},\ and\ \bibinfo
  {author} {\bibfnamefont {J.}~\bibnamefont {Richter}},\ }\href
  {https://doi.org/10.1103/PhysRevB.102.224303} {\bibfield  {journal} {\bibinfo
   {journal} {Phys. Rev. B}\ }\textbf {\bibinfo {volume} {102}},\ \bibinfo
  {pages} {224303} (\bibinfo {year} {2020})}\BibitemShut {NoStop}%
\bibitem [{\citenamefont {Vafek}\ \emph {et~al.}(2017)\citenamefont {Vafek},
  \citenamefont {Regnault},\ and\ \citenamefont
  {Bernevig}}]{vafek2017entanglement}%
  \BibitemOpen
  \bibfield  {author} {\bibinfo {author} {\bibfnamefont {O.}~\bibnamefont
  {Vafek}}, \bibinfo {author} {\bibfnamefont {N.}~\bibnamefont {Regnault}},\
  and\ \bibinfo {author} {\bibfnamefont {B.~A.}\ \bibnamefont {Bernevig}},\
  }\href {https://doi.org/doi: 10.21468/SciPostPhys.3.6.043} {\bibfield
  {journal} {\bibinfo  {journal} {SciPost Phys.}\ }\textbf {\bibinfo {volume}
  {3}},\ \bibinfo {pages} {043} (\bibinfo {year} {2017})}\BibitemShut {NoStop}%
\bibitem [{\citenamefont {Mark}\ and\ \citenamefont
  {Motrunich}(2020)}]{mark2020eta}%
  \BibitemOpen
  \bibfield  {author} {\bibinfo {author} {\bibfnamefont {D.~K.}\ \bibnamefont
  {Mark}}\ and\ \bibinfo {author} {\bibfnamefont {O.~I.}\ \bibnamefont
  {Motrunich}},\ }\href {https://doi.org/10.1103/PhysRevB.102.075132}
  {\bibfield  {journal} {\bibinfo  {journal} {Phys. Rev. B}\ }\textbf {\bibinfo
  {volume} {102}},\ \bibinfo {pages} {075132} (\bibinfo {year}
  {2020})}\BibitemShut {NoStop}%
\bibitem [{\citenamefont {Moudgalya}\ \emph
  {et~al.}(2020{\natexlab{c}})\citenamefont {Moudgalya}, \citenamefont
  {Regnault},\ and\ \citenamefont {Bernevig}}]{moudgalya2020eta}%
  \BibitemOpen
  \bibfield  {author} {\bibinfo {author} {\bibfnamefont {S.}~\bibnamefont
  {Moudgalya}}, \bibinfo {author} {\bibfnamefont {N.}~\bibnamefont
  {Regnault}},\ and\ \bibinfo {author} {\bibfnamefont {B.~A.}\ \bibnamefont
  {Bernevig}},\ }\href {https://doi.org/10.1103/PhysRevB.102.085140} {\bibfield
   {journal} {\bibinfo  {journal} {Phys. Rev. B}\ }\textbf {\bibinfo {volume}
  {102}},\ \bibinfo {pages} {085140} (\bibinfo {year}
  {2020}{\natexlab{c}})}\BibitemShut {NoStop}%
\bibitem [{\citenamefont {Hart}\ \emph {et~al.}(2020)\citenamefont {Hart},
  \citenamefont {De~Tomasi},\ and\ \citenamefont
  {Castelnovo}}]{hart2020compact}%
  \BibitemOpen
  \bibfield  {author} {\bibinfo {author} {\bibfnamefont {O.}~\bibnamefont
  {Hart}}, \bibinfo {author} {\bibfnamefont {G.}~\bibnamefont {De~Tomasi}},\
  and\ \bibinfo {author} {\bibfnamefont {C.}~\bibnamefont {Castelnovo}},\
  }\href {https://doi.org/10.1103/PhysRevResearch.2.043267} {\bibfield
  {journal} {\bibinfo  {journal} {Phys. Rev. Res.}\ }\textbf {\bibinfo {volume}
  {2}},\ \bibinfo {pages} {043267} (\bibinfo {year} {2020})}\BibitemShut
  {NoStop}%
\bibitem [{\citenamefont {Desaules}\ \emph {et~al.}(2021)\citenamefont
  {Desaules}, \citenamefont {Hudomal}, \citenamefont {Turner},\ and\
  \citenamefont {Papi{\'c}}}]{desaules2021proposal}%
  \BibitemOpen
  \bibfield  {author} {\bibinfo {author} {\bibfnamefont {J.-Y.}\ \bibnamefont
  {Desaules}}, \bibinfo {author} {\bibfnamefont {A.}~\bibnamefont {Hudomal}},
  \bibinfo {author} {\bibfnamefont {C.~J.}\ \bibnamefont {Turner}},\ and\
  \bibinfo {author} {\bibfnamefont {Z.}~\bibnamefont {Papi{\'c}}},\ }\href
  {https://doi.org/10.1103/PhysRevLett.126.210601} {\bibfield  {journal}
  {\bibinfo  {journal} {Phys. Rev. Lett.}\ }\textbf {\bibinfo {volume} {126}},\
  \bibinfo {pages} {210601} (\bibinfo {year} {2021})}\BibitemShut {NoStop}%
\bibitem [{\citenamefont {Pakrouski}\ \emph {et~al.}(2021)\citenamefont
  {Pakrouski}, \citenamefont {Pallegar}, \citenamefont {Popov},\ and\
  \citenamefont {Klebanov}}]{pakrouski2021group}%
  \BibitemOpen
  \bibfield  {author} {\bibinfo {author} {\bibfnamefont {K.}~\bibnamefont
  {Pakrouski}}, \bibinfo {author} {\bibfnamefont {P.~N.}\ \bibnamefont
  {Pallegar}}, \bibinfo {author} {\bibfnamefont {F.~K.}\ \bibnamefont
  {Popov}},\ and\ \bibinfo {author} {\bibfnamefont {I.~R.}\ \bibnamefont
  {Klebanov}},\ }\href {https://doi.org/10.1103/PhysRevResearch.3.043156}
  {\bibfield  {journal} {\bibinfo  {journal} {Phys. Rev. Res.}\ }\textbf
  {\bibinfo {volume} {3}},\ \bibinfo {pages} {043156} (\bibinfo {year}
  {2021})}\BibitemShut {NoStop}%
\bibitem [{\citenamefont {Yoshida}\ and\ \citenamefont
  {Katsura}(2022)}]{yoshida2022exact}%
  \BibitemOpen
  \bibfield  {author} {\bibinfo {author} {\bibfnamefont {H.}~\bibnamefont
  {Yoshida}}\ and\ \bibinfo {author} {\bibfnamefont {H.}~\bibnamefont
  {Katsura}},\ }\href {https://doi.org/10.1103/PhysRevB.105.024520} {\bibfield
  {journal} {\bibinfo  {journal} {Phys. Rev. B}\ }\textbf {\bibinfo {volume}
  {105}},\ \bibinfo {pages} {024520} (\bibinfo {year} {2022})}\BibitemShut
  {NoStop}%
\bibitem [{\citenamefont {Nakagawa}\ \emph {et~al.}(2022)\citenamefont
  {Nakagawa}, \citenamefont {Katsura},\ and\ \citenamefont
  {Ueda}}]{nakagawa2022exact}%
  \BibitemOpen
  \bibfield  {author} {\bibinfo {author} {\bibfnamefont {M.}~\bibnamefont
  {Nakagawa}}, \bibinfo {author} {\bibfnamefont {H.}~\bibnamefont {Katsura}},\
  and\ \bibinfo {author} {\bibfnamefont {M.}~\bibnamefont {Ueda}},\ }\href
  {https://doi.org/10.48550/arXiv.2205.07235} {\bibfield  {journal} {\bibinfo
  {journal} {arXiv:2205.07235}\ } (\bibinfo {year} {2022})}\BibitemShut
  {NoStop}%
\bibitem [{\citenamefont {Dutta}\ \emph {et~al.}(2015)\citenamefont {Dutta},
  \citenamefont {Gajda}, \citenamefont {Hauke}, \citenamefont {Lewenstein},
  \citenamefont {L{\"u}hmann}, \citenamefont {Malomed}, \citenamefont
  {Sowi{\'n}ski},\ and\ \citenamefont {Zakrzewski}}]{dutta2015non}%
  \BibitemOpen
  \bibfield  {author} {\bibinfo {author} {\bibfnamefont {O.}~\bibnamefont
  {Dutta}}, \bibinfo {author} {\bibfnamefont {M.}~\bibnamefont {Gajda}},
  \bibinfo {author} {\bibfnamefont {P.}~\bibnamefont {Hauke}}, \bibinfo
  {author} {\bibfnamefont {M.}~\bibnamefont {Lewenstein}}, \bibinfo {author}
  {\bibfnamefont {D.-S.}\ \bibnamefont {L{\"u}hmann}}, \bibinfo {author}
  {\bibfnamefont {B.~A.}\ \bibnamefont {Malomed}}, \bibinfo {author}
  {\bibfnamefont {T.}~\bibnamefont {Sowi{\'n}ski}},\ and\ \bibinfo {author}
  {\bibfnamefont {J.}~\bibnamefont {Zakrzewski}},\ }\href
  {https://iopscience.iop.org/article/10.1088/0034-4885/78/6/066001/meta}
  {\bibfield  {journal} {\bibinfo  {journal} {Rep. Prog. Phys.}\ }\textbf
  {\bibinfo {volume} {78}},\ \bibinfo {pages} {066001} (\bibinfo {year}
  {2015})}\BibitemShut {NoStop}%
\bibitem [{\citenamefont {Ruhman}\ and\ \citenamefont
  {Altman}(2017)}]{ruhman2017topological}%
  \BibitemOpen
  \bibfield  {author} {\bibinfo {author} {\bibfnamefont {J.}~\bibnamefont
  {Ruhman}}\ and\ \bibinfo {author} {\bibfnamefont {E.}~\bibnamefont
  {Altman}},\ }\href {https://doi.org/10.1103/PhysRevB.96.085133} {\bibfield
  {journal} {\bibinfo  {journal} {Phys. Rev. B}\ }\textbf {\bibinfo {volume}
  {96}},\ \bibinfo {pages} {085133} (\bibinfo {year} {2017})}\BibitemShut
  {NoStop}%
\bibitem [{\citenamefont {Gotta}\ \emph
  {et~al.}(2021{\natexlab{a}})\citenamefont {Gotta}, \citenamefont {Mazza},
  \citenamefont {Simon},\ and\ \citenamefont {Roux}}]{gotta2021two-prl}%
  \BibitemOpen
  \bibfield  {author} {\bibinfo {author} {\bibfnamefont {L.}~\bibnamefont
  {Gotta}}, \bibinfo {author} {\bibfnamefont {L.}~\bibnamefont {Mazza}},
  \bibinfo {author} {\bibfnamefont {P.}~\bibnamefont {Simon}},\ and\ \bibinfo
  {author} {\bibfnamefont {G.}~\bibnamefont {Roux}},\ }\href
  {https://doi.org/10.1103/PhysRevLett.126.206805} {\bibfield  {journal}
  {\bibinfo  {journal} {Phys. Rev. Lett.}\ }\textbf {\bibinfo {volume} {126}},\
  \bibinfo {pages} {206805} (\bibinfo {year} {2021}{\natexlab{a}})}\BibitemShut
  {NoStop}%
\bibitem [{\citenamefont {Gotta}\ \emph
  {et~al.}(2021{\natexlab{b}})\citenamefont {Gotta}, \citenamefont {Mazza},
  \citenamefont {Simon},\ and\ \citenamefont {Roux}}]{gotta2021two-prb}%
  \BibitemOpen
  \bibfield  {author} {\bibinfo {author} {\bibfnamefont {L.}~\bibnamefont
  {Gotta}}, \bibinfo {author} {\bibfnamefont {L.}~\bibnamefont {Mazza}},
  \bibinfo {author} {\bibfnamefont {P.}~\bibnamefont {Simon}},\ and\ \bibinfo
  {author} {\bibfnamefont {G.}~\bibnamefont {Roux}},\ }\href
  {https://doi.org/10.1103/PhysRevB.104.094521} {\bibfield  {journal} {\bibinfo
   {journal} {Phys. Rev. B}\ }\textbf {\bibinfo {volume} {104}},\ \bibinfo
  {pages} {094521} (\bibinfo {year} {2021}{\natexlab{b}})}\BibitemShut
  {NoStop}%
\bibitem [{\citenamefont {Di~Liberto}\ \emph {et~al.}(2014)\citenamefont
  {Di~Liberto}, \citenamefont {Creffield}, \citenamefont {Japaridze},\ and\
  \citenamefont {MoraisSmith}}]{di2014quantum}%
  \BibitemOpen
  \bibfield  {author} {\bibinfo {author} {\bibfnamefont {M.}~\bibnamefont
  {Di~Liberto}}, \bibinfo {author} {\bibfnamefont {C.~E.}\ \bibnamefont
  {Creffield}}, \bibinfo {author} {\bibfnamefont {G.~I.}\ \bibnamefont
  {Japaridze}},\ and\ \bibinfo {author} {\bibfnamefont {C.}~\bibnamefont
  {MoraisSmith}},\ }\href {https://doi.org/10.1103/PhysRevA.89.013624}
  {\bibfield  {journal} {\bibinfo  {journal} {Phys. Rev. A}\ }\textbf {\bibinfo
  {volume} {89}},\ \bibinfo {pages} {013624} (\bibinfo {year}
  {2014})}\BibitemShut {NoStop}%
\bibitem [{\citenamefont {Kohlert}\ \emph {et~al.}(2021)\citenamefont
  {Kohlert}, \citenamefont {Scherg}, \citenamefont {Sala}, \citenamefont
  {Pollmann}, \citenamefont {Madhusudhana}, \citenamefont {Bloch},\ and\
  \citenamefont {Aidelsburger}}]{kohlert2021experimental}%
  \BibitemOpen
  \bibfield  {author} {\bibinfo {author} {\bibfnamefont {T.}~\bibnamefont
  {Kohlert}}, \bibinfo {author} {\bibfnamefont {S.}~\bibnamefont {Scherg}},
  \bibinfo {author} {\bibfnamefont {P.}~\bibnamefont {Sala}}, \bibinfo {author}
  {\bibfnamefont {F.}~\bibnamefont {Pollmann}}, \bibinfo {author}
  {\bibfnamefont {B.~H.}\ \bibnamefont {Madhusudhana}}, \bibinfo {author}
  {\bibfnamefont {I.}~\bibnamefont {Bloch}},\ and\ \bibinfo {author}
  {\bibfnamefont {M.}~\bibnamefont {Aidelsburger}},\ }\href
  {https://doi.org/10.48550/arXiv.2106.15586} {\bibfield  {journal} {\bibinfo
  {journal} {arXiv:2106.15586}\ } (\bibinfo {year} {2021})}\BibitemShut
  {NoStop}%
\bibitem [{\citenamefont {Cr{\'e}pel}\ \emph {et~al.}(2022)\citenamefont
  {Cr{\'e}pel}, \citenamefont {Cea}, \citenamefont {Fu},\ and\ \citenamefont
  {Guinea}}]{crepel2022unconventional}%
  \BibitemOpen
  \bibfield  {author} {\bibinfo {author} {\bibfnamefont {V.}~\bibnamefont
  {Cr{\'e}pel}}, \bibinfo {author} {\bibfnamefont {T.}~\bibnamefont {Cea}},
  \bibinfo {author} {\bibfnamefont {L.}~\bibnamefont {Fu}},\ and\ \bibinfo
  {author} {\bibfnamefont {F.}~\bibnamefont {Guinea}},\ }\href
  {https://doi.org/10.1103/PhysRevB.105.094506} {\bibfield  {journal} {\bibinfo
   {journal} {Phys. Rev. B}\ }\textbf {\bibinfo {volume} {105}},\ \bibinfo
  {pages} {094506} (\bibinfo {year} {2022})}\BibitemShut {NoStop}%
\bibitem [{\citenamefont {Hudomal}\ \emph {et~al.}(2020)\citenamefont
  {Hudomal}, \citenamefont {Vasi{\'c}}, \citenamefont {Regnault},\ and\
  \citenamefont {Papi{\'c}}}]{hudomal2020quantum}%
  \BibitemOpen
  \bibfield  {author} {\bibinfo {author} {\bibfnamefont {A.}~\bibnamefont
  {Hudomal}}, \bibinfo {author} {\bibfnamefont {I.}~\bibnamefont {Vasi{\'c}}},
  \bibinfo {author} {\bibfnamefont {N.}~\bibnamefont {Regnault}},\ and\
  \bibinfo {author} {\bibfnamefont {Z.}~\bibnamefont {Papi{\'c}}},\ }\href
  {https://doi.org/10.1038/s42005-020-0364-9} {\bibfield  {journal} {\bibinfo
  {journal} {Commun. Phys.}\ }\textbf {\bibinfo {volume} {3}},\ \bibinfo
  {pages} {1} (\bibinfo {year} {2020})}\BibitemShut {NoStop}%
\bibitem [{\citenamefont {Zhao}\ \emph {et~al.}(2020)\citenamefont {Zhao},
  \citenamefont {Vovrosh}, \citenamefont {Mintert},\ and\ \citenamefont
  {Knolle}}]{zhao2020quantum}%
  \BibitemOpen
  \bibfield  {author} {\bibinfo {author} {\bibfnamefont {H.}~\bibnamefont
  {Zhao}}, \bibinfo {author} {\bibfnamefont {J.}~\bibnamefont {Vovrosh}},
  \bibinfo {author} {\bibfnamefont {F.}~\bibnamefont {Mintert}},\ and\ \bibinfo
  {author} {\bibfnamefont {J.}~\bibnamefont {Knolle}},\ }\href
  {https://doi.org/10.1103/PhysRevLett.124.160604} {\bibfield  {journal}
  {\bibinfo  {journal} {Phys. Rev. Lett.}\ }\textbf {\bibinfo {volume} {124}},\
  \bibinfo {pages} {160604} (\bibinfo {year} {2020})}\BibitemShut {NoStop}%
\bibitem [{\citenamefont {Zhao}\ \emph {et~al.}(2021)\citenamefont {Zhao},
  \citenamefont {Smith}, \citenamefont {Mintert},\ and\ \citenamefont
  {Knolle}}]{zhao2021orthogonal}%
  \BibitemOpen
  \bibfield  {author} {\bibinfo {author} {\bibfnamefont {H.}~\bibnamefont
  {Zhao}}, \bibinfo {author} {\bibfnamefont {A.}~\bibnamefont {Smith}},
  \bibinfo {author} {\bibfnamefont {F.}~\bibnamefont {Mintert}},\ and\ \bibinfo
  {author} {\bibfnamefont {J.}~\bibnamefont {Knolle}},\ }\href
  {https://doi.org/10.1103/PhysRevLett.127.150601} {\bibfield  {journal}
  {\bibinfo  {journal} {Phys. Rev. Lett.}\ }\textbf {\bibinfo {volume} {127}},\
  \bibinfo {pages} {150601} (\bibinfo {year} {2021})}\BibitemShut {NoStop}%
\bibitem [{\citenamefont {Shibata}\ \emph {et~al.}(2020)\citenamefont
  {Shibata}, \citenamefont {Yoshioka},\ and\ \citenamefont
  {Katsura}}]{shibata2020onsager}%
  \BibitemOpen
  \bibfield  {author} {\bibinfo {author} {\bibfnamefont {N.}~\bibnamefont
  {Shibata}}, \bibinfo {author} {\bibfnamefont {N.}~\bibnamefont {Yoshioka}},\
  and\ \bibinfo {author} {\bibfnamefont {H.}~\bibnamefont {Katsura}},\ }\href
  {https://doi.org/10.1103/PhysRevLett.124.180604} {\bibfield  {journal}
  {\bibinfo  {journal} {Phys. Rev. Lett.}\ }\textbf {\bibinfo {volume} {124}},\
  \bibinfo {pages} {180604} (\bibinfo {year} {2020})}\BibitemShut {NoStop}%
\bibitem [{\citenamefont {van Voorden}\ \emph {et~al.}(2021)\citenamefont {van
  Voorden}, \citenamefont {Marcuzzi}, \citenamefont {Schoutens},\ and\
  \citenamefont {Min{\'a}{\v{r}}}}]{van2021disorder}%
  \BibitemOpen
  \bibfield  {author} {\bibinfo {author} {\bibfnamefont {B.}~\bibnamefont {van
  Voorden}}, \bibinfo {author} {\bibfnamefont {M.}~\bibnamefont {Marcuzzi}},
  \bibinfo {author} {\bibfnamefont {K.}~\bibnamefont {Schoutens}},\ and\
  \bibinfo {author} {\bibfnamefont {J.}~\bibnamefont {Min{\'a}{\v{r}}}},\
  }\href {https://doi.org/10.1103/PhysRevB.103.L220301} {\bibfield  {journal}
  {\bibinfo  {journal} {Phys. Rev. B}\ }\textbf {\bibinfo {volume} {103}},\
  \bibinfo {pages} {L220301} (\bibinfo {year} {2021})}\BibitemShut {NoStop}%
\bibitem [{\citenamefont {Bariev}(1991)}]{bariev1991integrable}%
  \BibitemOpen
  \bibfield  {author} {\bibinfo {author} {\bibfnamefont {R.}~\bibnamefont
  {Bariev}},\ }\href {https://doi.org/10.1088/0305-4470/24/10/010} {\bibfield
  {journal} {\bibinfo  {journal} {J. Phys. A}\ }\textbf {\bibinfo {volume}
  {24}},\ \bibinfo {pages} {L549} (\bibinfo {year} {1991})}\BibitemShut
  {NoStop}%
\bibitem [{\citenamefont {Chhajlany}\ \emph {et~al.}(2016)\citenamefont
  {Chhajlany}, \citenamefont {Grzybowski}, \citenamefont {Stasi{\'n}ska},
  \citenamefont {Lewenstein},\ and\ \citenamefont
  {Dutta}}]{chhajlany2016hidden}%
  \BibitemOpen
  \bibfield  {author} {\bibinfo {author} {\bibfnamefont {R.~W.}\ \bibnamefont
  {Chhajlany}}, \bibinfo {author} {\bibfnamefont {P.~R.}\ \bibnamefont
  {Grzybowski}}, \bibinfo {author} {\bibfnamefont {J.}~\bibnamefont
  {Stasi{\'n}ska}}, \bibinfo {author} {\bibfnamefont {M.}~\bibnamefont
  {Lewenstein}},\ and\ \bibinfo {author} {\bibfnamefont {O.}~\bibnamefont
  {Dutta}},\ }\href {https://doi.org/10.1103/PhysRevLett.116.225303} {\bibfield
   {journal} {\bibinfo  {journal} {Phys. Rev. Lett.}\ }\textbf {\bibinfo
  {volume} {116}},\ \bibinfo {pages} {225303} (\bibinfo {year}
  {2016})}\BibitemShut {NoStop}%
\bibitem [{Note1()}]{Note1}%
  \BibitemOpen
  \bibinfo {note} {The term corresponding to $c_j^{(3)}$ gives rise to a
  nonlocal term that does not commute with the fermionic parity, rendering its
  fermionic counterpart unphysical.}\BibitemShut {Stop}%
\bibitem [{\citenamefont {Casati}\ \emph {et~al.}(1985)\citenamefont {Casati},
  \citenamefont {Chirikov},\ and\ \citenamefont {Guarneri}}]{casati1985energy}%
  \BibitemOpen
  \bibfield  {author} {\bibinfo {author} {\bibfnamefont {G.}~\bibnamefont
  {Casati}}, \bibinfo {author} {\bibfnamefont {B.~V.}\ \bibnamefont
  {Chirikov}},\ and\ \bibinfo {author} {\bibfnamefont {I.}~\bibnamefont
  {Guarneri}},\ }\href {https://doi.org/10.1103/PhysRevLett.54.1350} {\bibfield
   {journal} {\bibinfo  {journal} {Phys. Rev. Lett.}\ }\textbf {\bibinfo
  {volume} {54}},\ \bibinfo {pages} {1350} (\bibinfo {year}
  {1985})}\BibitemShut {NoStop}%
\bibitem [{\citenamefont {Prosen}\ and\ \citenamefont
  {Robnik}(1993)}]{prosen1993energy}%
  \BibitemOpen
  \bibfield  {author} {\bibinfo {author} {\bibfnamefont {T.}~\bibnamefont
  {Prosen}}\ and\ \bibinfo {author} {\bibfnamefont {M.}~\bibnamefont
  {Robnik}},\ }\href {https://doi.org/10.1088/0305-4470/26/10/010} {\bibfield
  {journal} {\bibinfo  {journal} {J. Phys. A}\ }\textbf {\bibinfo {volume}
  {26}},\ \bibinfo {pages} {2371} (\bibinfo {year} {1993})}\BibitemShut
  {NoStop}%
\bibitem [{\citenamefont {Pal}\ and\ \citenamefont {Huse}(2010)}]{pal2010many}%
  \BibitemOpen
  \bibfield  {author} {\bibinfo {author} {\bibfnamefont {A.}~\bibnamefont
  {Pal}}\ and\ \bibinfo {author} {\bibfnamefont {D.~A.}\ \bibnamefont {Huse}},\
  }\href {https://doi.org/10.1103/PhysRevB.82.174411} {\bibfield  {journal}
  {\bibinfo  {journal} {Phys. Rev. B}\ }\textbf {\bibinfo {volume} {82}},\
  \bibinfo {pages} {174411} (\bibinfo {year} {2010})}\BibitemShut {NoStop}%
\bibitem [{\citenamefont {Atas}\ \emph {et~al.}(2013)\citenamefont {Atas},
  \citenamefont {Bogomolny}, \citenamefont {Giraud},\ and\ \citenamefont
  {Roux}}]{atas2013distribution}%
  \BibitemOpen
  \bibfield  {author} {\bibinfo {author} {\bibfnamefont {Y.~Y.}\ \bibnamefont
  {Atas}}, \bibinfo {author} {\bibfnamefont {E.}~\bibnamefont {Bogomolny}},
  \bibinfo {author} {\bibfnamefont {O.}~\bibnamefont {Giraud}},\ and\ \bibinfo
  {author} {\bibfnamefont {G.}~\bibnamefont {Roux}},\ }\href
  {https://doi.org/10.1103/PhysRevLett.110.084101} {\bibfield  {journal}
  {\bibinfo  {journal} {Phys. Rev. Lett.}\ }\textbf {\bibinfo {volume} {110}},\
  \bibinfo {pages} {084101} (\bibinfo {year} {2013})}\BibitemShut {NoStop}%
\bibitem [{Note2()}]{Note2}%
  \BibitemOpen
  \bibinfo {note} {Here and throughout the paper, $|{\protect \cal S}|$ denotes
  the number of elements in a set ${\protect \cal S}$.}\BibitemShut {Stop}%
\bibitem [{\citenamefont {Yang}(1989)}]{yang1989eta}%
  \BibitemOpen
  \bibfield  {author} {\bibinfo {author} {\bibfnamefont {C.~N.}\ \bibnamefont
  {Yang}},\ }\href {https://doi.org/10.1103/PhysRevLett.63.2144} {\bibfield
  {journal} {\bibinfo  {journal} {Phys. Rev. Lett.}\ }\textbf {\bibinfo
  {volume} {63}},\ \bibinfo {pages} {2144} (\bibinfo {year}
  {1989})}\BibitemShut {NoStop}%
\bibitem [{\citenamefont {Gotta}\ \emph {et~al.}(2022)\citenamefont {Gotta},
  \citenamefont {Mazza}, \citenamefont {Simon},\ and\ \citenamefont
  {Roux}}]{gotta2022exact}%
  \BibitemOpen
  \bibfield  {author} {\bibinfo {author} {\bibfnamefont {L.}~\bibnamefont
  {Gotta}}, \bibinfo {author} {\bibfnamefont {L.}~\bibnamefont {Mazza}},
  \bibinfo {author} {\bibfnamefont {P.}~\bibnamefont {Simon}},\ and\ \bibinfo
  {author} {\bibfnamefont {G.}~\bibnamefont {Roux}},\ }\href
  {https://doi.org/10.48550/arXiv.2207.07531} {\bibfield  {journal} {\bibinfo
  {journal} {arXiv:2207.07531}\ } (\bibinfo {year} {2022})}\BibitemShut
  {NoStop}%
\bibitem [{\citenamefont {Bu{\v{c}}a}\ \emph {et~al.}(2019)\citenamefont
  {Bu{\v{c}}a}, \citenamefont {Tindall},\ and\ \citenamefont
  {Jaksch}}]{buvca2019non}%
  \BibitemOpen
  \bibfield  {author} {\bibinfo {author} {\bibfnamefont {B.}~\bibnamefont
  {Bu{\v{c}}a}}, \bibinfo {author} {\bibfnamefont {J.}~\bibnamefont
  {Tindall}},\ and\ \bibinfo {author} {\bibfnamefont {D.}~\bibnamefont
  {Jaksch}},\ }\href {https://doi.org/10.1038/s41467-019-09757-y} {\bibfield
  {journal} {\bibinfo  {journal} {Nat. Commun.}\ }\textbf {\bibinfo {volume}
  {10}},\ \bibinfo {pages} {1} (\bibinfo {year} {2019})}\BibitemShut {NoStop}%
\bibitem [{\citenamefont {Mori}\ \emph {et~al.}(2018)\citenamefont {Mori},
  \citenamefont {Ikeda}, \citenamefont {Kaminishi},\ and\ \citenamefont
  {Ueda}}]{mori2018thermalization}%
  \BibitemOpen
  \bibfield  {author} {\bibinfo {author} {\bibfnamefont {T.}~\bibnamefont
  {Mori}}, \bibinfo {author} {\bibfnamefont {T.~N.}\ \bibnamefont {Ikeda}},
  \bibinfo {author} {\bibfnamefont {E.}~\bibnamefont {Kaminishi}},\ and\
  \bibinfo {author} {\bibfnamefont {M.}~\bibnamefont {Ueda}},\ }\href
  {https://doi.org/10.1088/1361-6455/aabcdf} {\bibfield  {journal} {\bibinfo
  {journal} {J. Phys. B}\ }\textbf {\bibinfo {volume} {51}},\ \bibinfo {pages}
  {112001} (\bibinfo {year} {2018})}\BibitemShut {NoStop}%
\bibitem [{\citenamefont {Vidmar}\ and\ \citenamefont
  {Rigol}(2017)}]{vidmar2017entanglement}%
  \BibitemOpen
  \bibfield  {author} {\bibinfo {author} {\bibfnamefont {L.}~\bibnamefont
  {Vidmar}}\ and\ \bibinfo {author} {\bibfnamefont {M.}~\bibnamefont {Rigol}},\
  }\href {https://doi.org/https://doi.org/10.1103/PhysRevLett.119.220603}
  {\bibfield  {journal} {\bibinfo  {journal} {Phys. Rev. Lett.}\ }\textbf
  {\bibinfo {volume} {119}},\ \bibinfo {pages} {220603} (\bibinfo {year}
  {2017})}\BibitemShut {NoStop}%
\bibitem [{\citenamefont {Bianchi}\ and\ \citenamefont
  {Dona}(2019)}]{bianchi2019typical}%
  \BibitemOpen
  \bibfield  {author} {\bibinfo {author} {\bibfnamefont {E.}~\bibnamefont
  {Bianchi}}\ and\ \bibinfo {author} {\bibfnamefont {P.}~\bibnamefont {Dona}},\
  }\href {https://doi.org/https://doi.org/10.1103/PhysRevD.100.105010}
  {\bibfield  {journal} {\bibinfo  {journal} {Phys, Rev, D}\ }\textbf {\bibinfo
  {volume} {100}},\ \bibinfo {pages} {105010} (\bibinfo {year}
  {2019})}\BibitemShut {NoStop}%
\bibitem [{\citenamefont {Bianchi}\ \emph {et~al.}(2022)\citenamefont
  {Bianchi}, \citenamefont {Hackl}, \citenamefont {Kieburg}, \citenamefont
  {Rigol},\ and\ \citenamefont {Vidmar}}]{bianchi2022volume}%
  \BibitemOpen
  \bibfield  {author} {\bibinfo {author} {\bibfnamefont {E.}~\bibnamefont
  {Bianchi}}, \bibinfo {author} {\bibfnamefont {L.}~\bibnamefont {Hackl}},
  \bibinfo {author} {\bibfnamefont {M.}~\bibnamefont {Kieburg}}, \bibinfo
  {author} {\bibfnamefont {M.}~\bibnamefont {Rigol}},\ and\ \bibinfo {author}
  {\bibfnamefont {L.}~\bibnamefont {Vidmar}},\ }\href
  {https://doi.org/10.1103/PRXQuantum.3.030201} {\bibfield  {journal} {\bibinfo
   {journal} {PRX Quantum}\ }\textbf {\bibinfo {volume} {3}},\ \bibinfo {pages}
  {030201} (\bibinfo {year} {2022})}\BibitemShut {NoStop}%
\bibitem [{\citenamefont {Tanaka}(2008)}]{tanaka2008ferromagnetic}%
  \BibitemOpen
  \bibfield  {author} {\bibinfo {author} {\bibfnamefont {A.}~\bibnamefont
  {Tanaka}},\ }\href {https://doi.org/10.1088/1751-8113/41/36/365208}
  {\bibfield  {journal} {\bibinfo  {journal} {J. Phys. A}\ }\textbf {\bibinfo
  {volume} {41}},\ \bibinfo {pages} {365208} (\bibinfo {year}
  {2008})}\BibitemShut {NoStop}%
\bibitem [{\citenamefont {Weinberg}\ and\ \citenamefont
  {Bukov}(2017)}]{weinberg2017quspin}%
  \BibitemOpen
  \bibfield  {author} {\bibinfo {author} {\bibfnamefont {P.}~\bibnamefont
  {Weinberg}}\ and\ \bibinfo {author} {\bibfnamefont {M.}~\bibnamefont
  {Bukov}},\ }\href {https://doi.org/10.21468/SciPostPhys.2.1.003} {\bibfield
  {journal} {\bibinfo  {journal} {SciPost Phys.}\ }\textbf {\bibinfo {volume}
  {2}},\ \bibinfo {pages} {003} (\bibinfo {year} {2017})}\BibitemShut {NoStop}%
\bibitem [{\citenamefont {Weinberg}\ and\ \citenamefont
  {Bukov}(2019)}]{weinberg2019quspin}%
  \BibitemOpen
  \bibfield  {author} {\bibinfo {author} {\bibfnamefont {P.}~\bibnamefont
  {Weinberg}}\ and\ \bibinfo {author} {\bibfnamefont {M.}~\bibnamefont
  {Bukov}},\ }\href {https://doi.org/10.21468/SciPostPhys.7.2.020} {\bibfield
  {journal} {\bibinfo  {journal} {SciPost Phys.}\ }\textbf {\bibinfo {volume}
  {7}},\ \bibinfo {pages} {020} (\bibinfo {year} {2019})}\BibitemShut {NoStop}%
\end{thebibliography}%

\end{document}